\documentclass{article}
\pdfoutput=1
\title{Respondent-Driven Sampling:\\  An Assessment of Current Methodology}

\author{Krista J. Gile and Mark S. Handcock}

\usepackage{amssymb,amsmath,amsthm}
\usepackage{graphicx}
\usepackage{mathrsfs} 
\usepackage{natbib}
\usepackage{amsmath}
\usepackage{amssymb}
\usepackage[all]{xy}
\usepackage{color, graphics}
\usepackage{palatino, url, multicol}
\usepackage{graphicx}
\usepackage{subfigure}
\usepackage{mathrsfs}
\usepackage{multicol}
\newcommand{\bi}{\begin{itemize}}
\newcommand{\ei}{\end{itemize}}

\newcommand{\IR}{\mathbb{R}}

\definecolor{Emphcolor}{cmyk}{0,0.89,0.94,0.1}
\definecolor{Netcolor}{rgb}{.8,0,.9}
\definecolor{Diseasecolor}{rgb}{1,.8,.2}
\definecolor{Sampcolor}{rgb}{0,.9,.3}
\definecolor{Black}{rgb}{0,0,0}
\definecolor{Red}{rgb}{1,0,0}
\definecolor{Blue}{rgb}{0,0,1}
\definecolor{Gray}{gray}{.6}

\newcommand{\red}{\color{Red}}
\newcommand{\blue}{\color{Blue}}

\font\elevenrm=cmr11
\newcommand{\ql}{{\elevenrm ``}}
\newcommand{\qr}{{\elevenrm "}\ }

\theoremstyle{plain}

\theoremstyle{definition}

\newcommand{\ecol}{\blue}
\newcommand{\eone}{}
\newcommand{\etwo}{}

\newcommand{\sss}{2.1in}
\newcommand{\ttt}{2in}

\usepackage{graphicx}
\usepackage{float}
\usepackage{rotating}
\usepackage{natbib}
\usepackage{color}
\usepackage{pdflscape}

\usepackage{graphics}
\usepackage{subfigure}
\usepackage[all]{xy}
\usepackage{amsmath}
\usepackage{amssymb}
\usepackage[all]{xy}
\usepackage{palatino, url, multicol}
\usepackage{mathrsfs}
\usepackage{multicol}
\usepackage{enumerate}

\definecolor{Emphcolor}{rgb}{.1,.1,.5}
\definecolor{Red}{rgb}{.9,0,.1}
\definecolor{Blue}{rgb}{.1,.1,.5}

\newcommand{\bea}{\begin{eqnarray}}
\newcommand{\eea}{\end{eqnarray}}

\usepackage{ifthen}
\newboolean{draft}
\setboolean{draft}{false}
\newcommand{\knote}[1]{\ifthenelse{\boolean{draft}}{{\bf knote:~}{\it
#1}\relax}{}}
\newcommand{\mnote}[1]{\ifthenelse{\boolean{draft}}{{\bf mnote:~}{\it
#1}\relax}{}}

\begin{document}
\maketitle

\renewenvironment{quotation} {\relax}

\begin{abstract}
Respondent-Driven Sampling (RDS) employs a variant of a link-tracing network sampling
strategy to collect data from hard-to-reach populations. By tracing the links
in the underlying social network, the process exploits the social structure to
expand the sample and reduce its dependence on the initial (convenience)
sample.

The primary goal of RDS is typically to estimate population averages in the hard-to-reach population.  The current estimates make strong assumptions in order to treat the data as a probability sample.  In particular, we evaluate three critical sensitivities of the estimators:  to bias induced by the initial sample, to uncontrollable features of respondent behavior, and to the without-replacement structure of sampling.

First, we address the reduction of bias induced by the initial convenience sample.  RDS relies on many sample waves to create a type of mixing in the sampling process, much like the mixing in a Markov chain.  We illustrate that the number of sample waves typically used in RDS is likely insufficient for the type of nodal mixing required to obtain the reputed asymptotic unbiasedness of the estimators.  Nevertheless, in some cases we find that the resulting estimators to be approximately unbiased, although this is highly sensitive to the degree of clustering in the population and the number of waves in the sample.

Second, we highlight the dependence of the estimators on characteristics of respondent behavior outside the control of the researcher.  In particular, we illustrate the bias induced in the estimator by preferential referral behavior by respondents.  We highlight the need to expand data collection to learn more about how respondents behave in RDS studies.

Finally, estimates are based on a with-replacement random walk model, while the actual sampling is without replacement.  We illustrate that when a substantial fraction
of the target population is sampled this approximation can lead to substantial bias in the resulting estimators.  Previous research on the properties of RDS estimators has ignored the without-replacement nature of the sample.

This paper sounds a cautionary note for the users of RDS.  While
current RDS methodology is powerful and clever, the favorable
statistical properties claimed for the current estimates are
shown to be heavily dependent on often unrealistic assumptions.
\end{abstract}

\renewenvironment{quotation}
               {\list{}{\listparindent 1.5em%
                        \itemindent    \listparindent
                        \rightmargin   \leftmargin
                        \parsep        \z@ \@plus\p@}%
                \item\relax}
               {\endlist}


\vfil\eject
\setcounter{page}{1}

\section{Introduction to Respondent-Driven Sampling}
Respondent-Driven Sampling (RDS, introduced by Heckathorn 1997, 2002, see also Salganik and Heckathorn, 2004, Volz and Heckathorn 2008\nocite{heck97, Heckathorn2002, salgheck04, volzheck08}) is an approach to sampling design and inference in hard-to-reach populations.  {\it Hard-to-reach} populations are characterized by the difficulty in sampling from them using standard probability methods.  RDS is typically employed when a sampling frame for the target population is not available, and its members are rare or stigmatized in the larger population so that it is prohibitively expensive to contact them through available frames.   It is often used in populations such as injection drug users, men who have sex with men, and sex workers \citep{Malekinejad:2008ty}, although it has also been used in other populations such as jazz musicians \citep{JazzMusicians2001}, unregulated workers \citep{bernhardt2006}, and native American subgroups \citep{walters2002}.

RDS presents two main innovations for this setting:  a design for sampling from the target population and a corresponding strategy for estimating properties of the target population based on the resulting sample.  It is from the former that the method draws its name:  the {\it Respondent-Driven} sampling design relies on the respondents at each wave to select or {\ql}drive\qr the next wave of sampling through their selection of other members of the target population.  This is typically achieved through the distribution of coupons by respondents to other members of the target population.  Thus, RDS exploits the network of social relations connecting the target population to facilitate sampling.  This strategy also reduces the confidentiality concerns generally associated with sampling from stigmatized populations.

The second main innovation is in estimating population characteristics based on the sample.
As with most studies of hidden populations, a RDS sample begins with a convenience sample of individuals.  The key innovation is that through many waves of sampling, the dependence of the final sample on the initial convenience sample is reduced.  The estimates of inclusion probabilities in current RDS inference rely on arguments based on a Markov Chain representation of the sampling process.  
This innovation was proposed by \cite{salgheck04} and extended by \cite{volzheck08}.

	RDS employs a {\it link-tracing} sampling design.  In such designs, network links from sampled members of the target population are followed (traced) to select subsequent population members to add to the sample.  In the case of RDS, the network links of interest are the social contacts facilitating the transfer of RDS coupons.  Two population members related by such a link are said to be {\it alters} of one another.
In the context of hard-to-reach populations such strategies are often referred to as {\it snowball} samples \citep{goodman1961}.  Snowball sampling is useful in settings where a network of social relations links the members of the target population, such that previously sampled individuals can facilitate the sampling of others in the population.
Such samples are often very effective at recruiting large samples from hard-to-reach populations.  Despite Goodman's probabilistic formulation, however, the initial sample is typically a convenience sample, such that the ultimate snowball sample is not a probability sample (i.e. the probabilities of samples are not computable).  Therefore, in most snowball samples from hard-to-reach populations, valid statistical inference is not forthcoming.
	
	In RDS, the initial sample (also know as the {\it seeds} or {\it $0^{th}$ wave}) is assumed to be a convenience sample, selected from among the members of the target population known to the researchers.  Each respondent is then given a fixed small number of coupons to distribute among their alters in the target population.  Each successive {\it wave} of the sample consists of population members who are given coupons by members of the previous wave and return those coupons to the survey center.  A respondent typically receives additional compensation for each successful recruitment.  Respondents are also asked to report their numbers of contacts within the target population, to be used as an estimate of their nodal {\it degree} or number of alters.  The passing of coupons reduces confidentiality concerns in marginalized populations, and the dual incentive structure encourages the buy-in of participant-recruiters.  The limited number of coupons and measurement of degree facilitate the estimation approach described in Section \ref{sec:rdsest}.

\subsection{RDS Addresses an Under-Served Need}
Absent respondent-driven sampling, frameworks for gathering probability samples of hard-to-reach populations are few and unappealing.  A time-location sample \citep{muhib01,peterson_etal2008} will generate a probability sample, but with probabilities conditional on times and locations, rather than population members.
A probability sample from a larger frame such as a door-to-door survey may generate a probability sample, but the rarity of the target population may make such a procedure prohibitively expensive.  This type of study would also need to negotiate the difficult task of soliciting potentially sensitive information about membership in a marginalized population.  It is also possible members of the marginalized population would be under-represented in a standard sampling frame.  Among non-probability sampling methods, targeted sampling \citep{targetedsampling1989} is among the most promising.  This approach combines extensive foundational research with a flexible form of quota sampling to improve the breadth of the sample.  The resulting sample, however, is not a probability sample, and one study \citep{aqcdc06} finds the resulting sample is less diverse than that of a parallel sample collected through RDS.

The need for RDS is demonstrated by the recent explosion of RDS studies, both in the US and abroad.  \cite{johnston08} cite 128 current or completed RDS studies from 30 countries outside the US.  These studies have taken place on several continents (Europe, Asia, South America, Africa, and Australia), and focused primarily on populations of  injection drug users, men who have sex with men,  and sex workers.

Notable among the multitude of RDS studies in the U.S. is the recent use of RDS for behavioral monitoring by the Centers for Disease Control and Prevention (CDC). \cite{aqcdc06} describe this study, in which the CDC is using RDS for behavioral surveillance of high-risk HIV-related behaviors among injection drug users.  The overall study is called the National HIV Behavioral Surveillance System (NHBS), and consists of rotating studies in three high-risk populations:  Men who have sex with men (MSM:  NHBS-MSM), injection drug users (IDU:  NHBS-IDU), and high-risk heterosexuals (NHBS-HET).  The study is being conducted in 25 metropolitan statistical areas (MSAs), each of which conducts a study of one of these sub-populations each year, on a rotating annual schedule.
The study is interested in informing policies related to behavior change.  Therefore, the primary scientific questions of this study concern risk behaviors and treatments.   The simulation study presented in this paper is based as closely as possible on replicating the conditions described in the pilot study data in \cite{aqcdc06}.

\subsection{Existing Literature Analyzing RDS Estimators}
In this paper, we use a simulation study to assess the performance of existing RDS estimators in three main areas:
\begin{enumerate}
  \item Sensitivity to the procedure for selecting the seeds.
  \item Sensitivity to respondent behavior.
  \item Sensitivity to the with-replacement sampling assumption.
\end{enumerate}
Other studies have addressed sensitivity to seed selection and, to a lesser degree, respondent behavior, but this paper is the first to conduct such a study in the context of without-replacement sampling, and is also the first to address without-replacement sampling directly.  This is partially because the without-replacement nature of the sampling process is difficult to capture analytically.  The difficulty is exacerbated by the {\it branching} structure of the RDS sample, that is the fact that each respondent can recruit multiple alters.




Others have applied analytical analyses to evaluating RDS, but without treating the with-replacement or branching features of RDS.  \cite{Goel2007} are the first to attempt this, using a Markov chain model.  They analytically demonstrate that increased variance is associated with elevated levels of clustering in the underlying population.  They then use a simulation study to illustrate the increased variance resulting from the branching structure of the sample.  \cite{neely09} considers a Bayesian frame, in which he assigns parameters to many features of the with-replacement model of RDS sampling.  Based on this model, he allows for prior selection to represent the researcher's degree of confidence in two aspects of respondent behavior: randomness of recruitment from among alters and accurate degree reporting, and illustrates the large amounts of uncertainty thus introduced into the estimator.  This analysis does not address aspects of seed selection or consider with-replacement or branching sampling.

 Previous simulation studies have been small and mostly embedded in the introduction of new estimators.  In the introduction of their new estimator, \cite{salgheck04} include a simulation illustrating the rapid convergence of their estimator to the true population proportion.  They focus on the convergence of the expected value in a with-replacement setting with seed selection procedures which vary from proportional to degree (the equilibrium of the process they consider) to completely at random.  Our simulations (not highlighted in this paper) show that these two extremes of seed selection produce little difference in the resulting estimators, although we find strong effects from more strongly biased seed selection regimes.  \cite{volzheck08} also include a simulation study, in which they illustrate that the uncertainty of the estimates decreases with increasing sample size, and increases with stronger clustering, again in a without-replacement non-branching frame.  \cite{Goel2007} present a simulation study examining the impact of the branching structure of RDS. They illustrate that a branching with-replacement random walk sample results in a much more variable estimator than a non-branching with-replacement random walk sample of the same sample size.

\cite{wejnertheck08} introduce a novel and important means of evaluating RDS estimation.  They apply the RDS sampling procedure to a population with known characteristics: undergraduates in a large residential university.  They execute their sample via e-mail referral and web surveys, and collect a total of 159 samples from a population of about 13,000.  They find that most of their confidence intervals do include the true parameter values, although they do find evidence of biased referral of alters.  While this is helpful, it is likely that the behavior of undergraduates sending e-mails might differ in critical ways from the behavior of other populations such as injection drug users passing coupons.

Overall, then, RDS estimators are overdue for a systematic evaluation considering sensitivity to seed selection, respondent behavior, and without-replacement, sampling.  This paper aims to fill this need.

\subsection{Outline of the Paper}

The properties of existing estimators for RDS data are not well understood.
The theoretical model motivating the estimation is known to be an approximation.  Even under this theoretical model, the estimated sampling probabilities are known to be inaccurate for at least the early waves of the sample.   Many of the other assumptions have not been explored systematically.

RDS data collection is a complicated procedure involving a branching without-replacement process, on an arbitrary graph of social relations, starting from a convenience sample of seeds.  It is a very complex stochastic process.  We considered several analytical approximations to this process, but found them to be either so complex as to be intractable, or unable to model the critical branching and without-replacement features of the sampling design \citep{gile08}.  For this reason, the present evaluation relies on a simulation study.

In this study, we use computational simulations to systematically study several dimensions of the RDS sampling process, retaining the critical without-replacement and branching features ignored in most previous studies.


In Section \ref{sec:rdsest}, we review the existing estimators used for RDS data and
introduce a framework to understand and assess the assumptions that
underly them.
In Section \ref{section:simstudy}, we use a simulation study to evaluate the properties of the RDS estimator proposed in \cite{volzheck08} under conditions of violations of some of the assumptions.
We evaluate the classic RDS estimator proposed by \cite{salgheck04} in Section \ref{sec:sh}, in the interest in comparing its performance to that of the Volz-Heckathorn estimator.  In Section \ref{sec:ch5discuss} we summarize our assessment of current RDS methodology, highlighting areas for future work.

\section{Overview of Existing RDS Estimation} \label{sec:rdsest}
The basic ideas underlying estimation from RDS data are clever and important. They motivate estimators for hidden population characteristics that are elusive based on alternative sampling methods.  They allow for some {\it something like} valid statistical inference, in a sampling setting where the target population cannot be reached in a systematic manner.  Unfortunately, the existing literature sometimes understates the degree of dependence on underlying
assumptions, and the tenuous nature of some of those assumptions.

The original article, \citep{heck97} made very strong assumptions about the sampling procedure so as to assume that the sample proportions were representative of the population proportions.  \cite{salgheck04} introduced a Markov chain argument for population mixing, and proposed an estimator based on equating the number of cross-relations between pairs of sub-populations of interest, based on the referral patterns of each group.  This estimator is currently in wide use, and is implemented in the standard RDS analysis software \citep{rdsat}.  We refer to this estimator as the {\it classic} RDS estimator, or as the Salganik-Heckathorn (S-H) estimator.  These earlier papers made two critical contributions.  First, they introduced a sampling strategy that is practically feasible in a large number hard-to-reach populations.  Second, they introduced the key insight of leveraging many waves of sampling to reduce dependence on the initial convenience sample to increase the validity of statistical inference.

\cite{volzheck08} connect RDS estimation to mainstream survey sampling through the use of a generalized Horvitz-Thompson estimator form.  This estimator relies heavily on the estimation of the inclusion probabilities of the sampled units, $\pi_i$.  We refer to this approach as the {\it current} RDS estimator, or as the Volz-Heckathorn (V-H) estimator.  We expect the V-H estimator to supplant the S-H estimator in common usage, perhaps during the publication process of this paper.  We also find that the V-H estimator out-performs the S-H estimator in almost all circumstances (Section \ref{sec:sh}).
For these reasons, most of the results in this paper pertain directly to the V-H estimator.  The exception is Section \ref{sec:sh}, in which we present a comparison of the V-H and S-H estimators, and describe the differences in their performances across our simulation studies.

\subsection{The Current RDS Estimator}\label{sec:estimator}
The V-H estimator is a variant of the generalized Horvitz-Thompson estimator, using estimated inclusion probabilities and samples drawn using RDS.  We begin by introducing this more general estimator.

Consider a population of $N$ individuals. If we knew the probability, $\pi_i$ of including each sampled individual $i$ in the sample, we could estimate the population mean $\mu = \frac{1}{N}\sum_{i \in 1\ldots N}z_i$ of any quantity $z_i$ measured on the sampled individuals using a Horvitz-Thompson \citep{ht52} estimator:
\bea
\hat{\mu} = \frac{1}{N}\sum_{i: S_{i}=1}\frac{z_i}{\pi_i},
\eea
where $S$ is the random $N$-vector representing the sample, such that $S_i=1$ if unit $i$ is sampled, and is otherwise $0$.
There are two difficulties with this approach in the context of RDS:  the population size $N$ is unknown, and the inclusion probabilities $\pi_i$ are unknown.  The first is easily dispensed with.  Normalizing by an unbiased estimator of $N$, $\hat{N} = \sum_{i: S_{i}=1}\frac{1}{\pi_i}$, we obtain:
\bea
{\hat\mu}^* = \frac{\sum_{i: S_{i}=1}\frac{z_i}{\pi_i}}{\sum_{i: S_{i}=1}\frac{1}{\pi_i}},
\label{rdsht}
\eea
which is the ratio of two unbiased estimators and therefore tends to estimate $\mu$ with small bias for large sample sizes. 
This is a standard variant on the classic Horvitz-Thompson estimator, referred to as the {\it generalized Horvitz-Thompson estimator}.  \cite{thompson2002} suggests that this estimator typically out-performs the standard Horvitz-Thompson estimator when inclusion probabilities are far from  proportional to $z_i$.
More difficult is the estimation of the $\pi_i$.  RDS sampling design is focused on collecting a sizable sample of individuals, while allowing for the estimation of the inclusion probabilities $\pi_i$.

The V-H estimator relies on an approximation to the $\pi_i$ based on treating the sampling process as a random walk on the network connecting the target population.  Consider again a population of $N$ individuals or, structurally, {\it nodes}.
Let the $N \times N$ matrix $y$ represent the {\it sociomatrix} of relations in the population, such that $y_{ij}=1$ if $j$ is an alter of $i$
and $y_{ij}=0$ otherwise.  In keeping with the standard RDS assumption, we assume that $y$ is undirected, such that $y_{ij}=y_{ji}$.  As above, let the N-vector $z$ represent the nodal attribute of scientific interest.

Consider a random walk process on the graph given by
$y$.  This process is defined as follows:   Let the vector $G$ represent the successive indices of nodes sampled by the random walk process, such that $G_k$ is the index of the node sampled at the $k^{th}$ step.  Then the random walk process is a Markov process on the space of nodal indices characterized by the transition matrix:

\begin{equation}
P(G_{k+1}=j | G_k = i) = T_{ij} = \left\{ \begin{array}{rl}
\frac{1}{d_i} & y_{ij} = 1\\
0 & y_{ij} = 0
\end{array}\right.
\end{equation}
where $d_i = \sum_j y_{ij}$ is the degree of node $i$.  This constitutes (completely at) {\it random referral} among alters.  Then if the graph $y$ is a single connected component or {\it connected graph}, this process constitutes an irreducible Markov Chain on the space of the nodal indices, characterized by transition matrix $T$ with unique stationary distribution given by draw-wise selection probabilities $p_i$ proportional to degree, or $p_i = \alpha d_i$ for some constant $\alpha$.

\begin{figure}[h]
\begin{center}
    \includegraphics[width=4in]{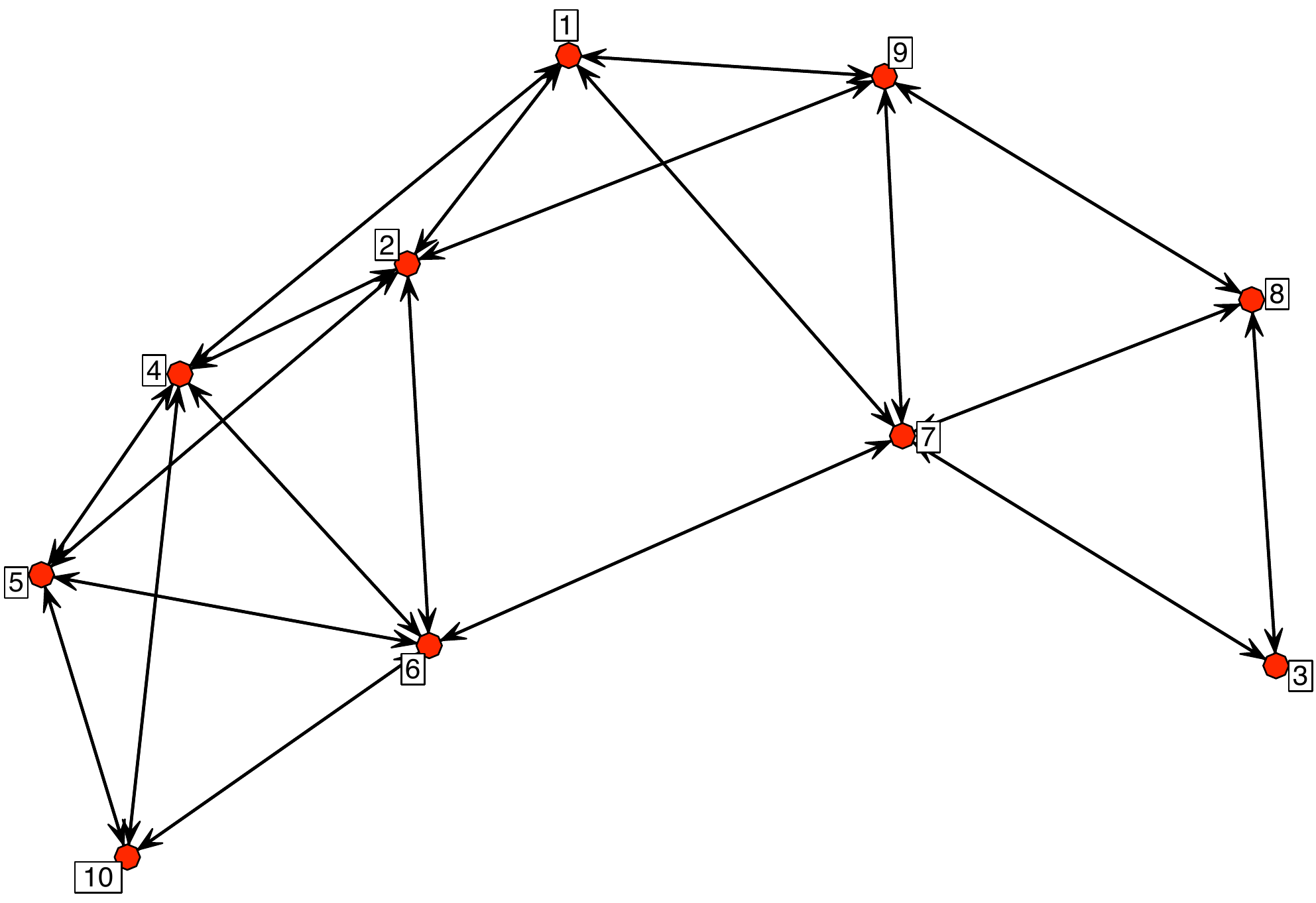}
\end{center}
\caption{Network used to illustrate convergence of sampling probabilities} \label{sociogram0228b}
\end{figure}

\begin{figure}[h]
\begin{center}
    \includegraphics[width=2in,height=2in]{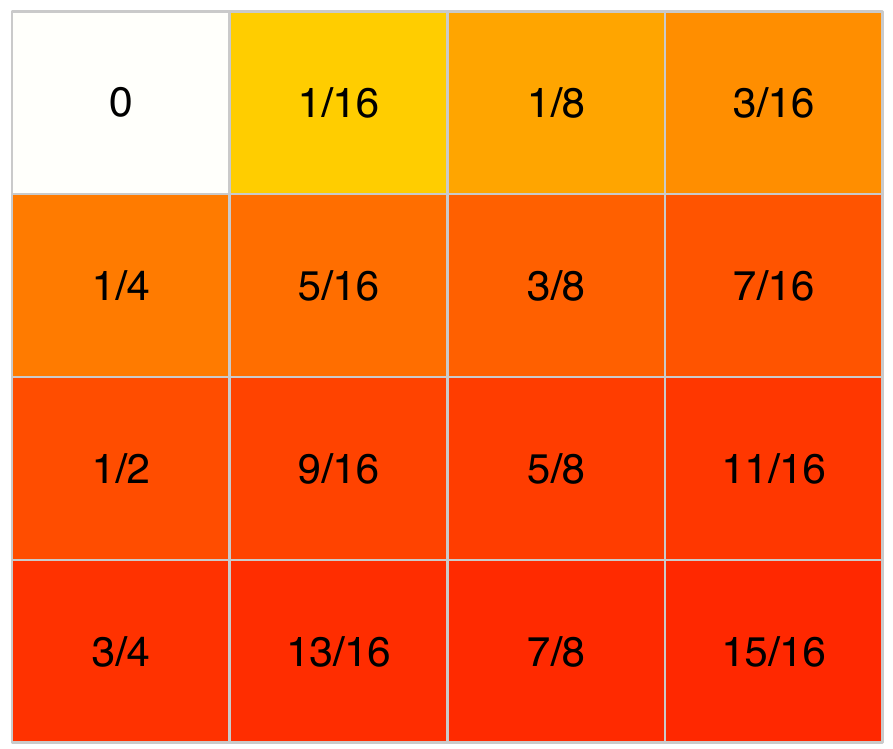}
\end{center}
\caption[Color Mapping of Probabilities]{Legend for the color mapping of probability space} \label{cdcheatprobs}
\end{figure}

\begin{figure}[h]
\begin{center}
    \includegraphics[width=5in,height=5in]{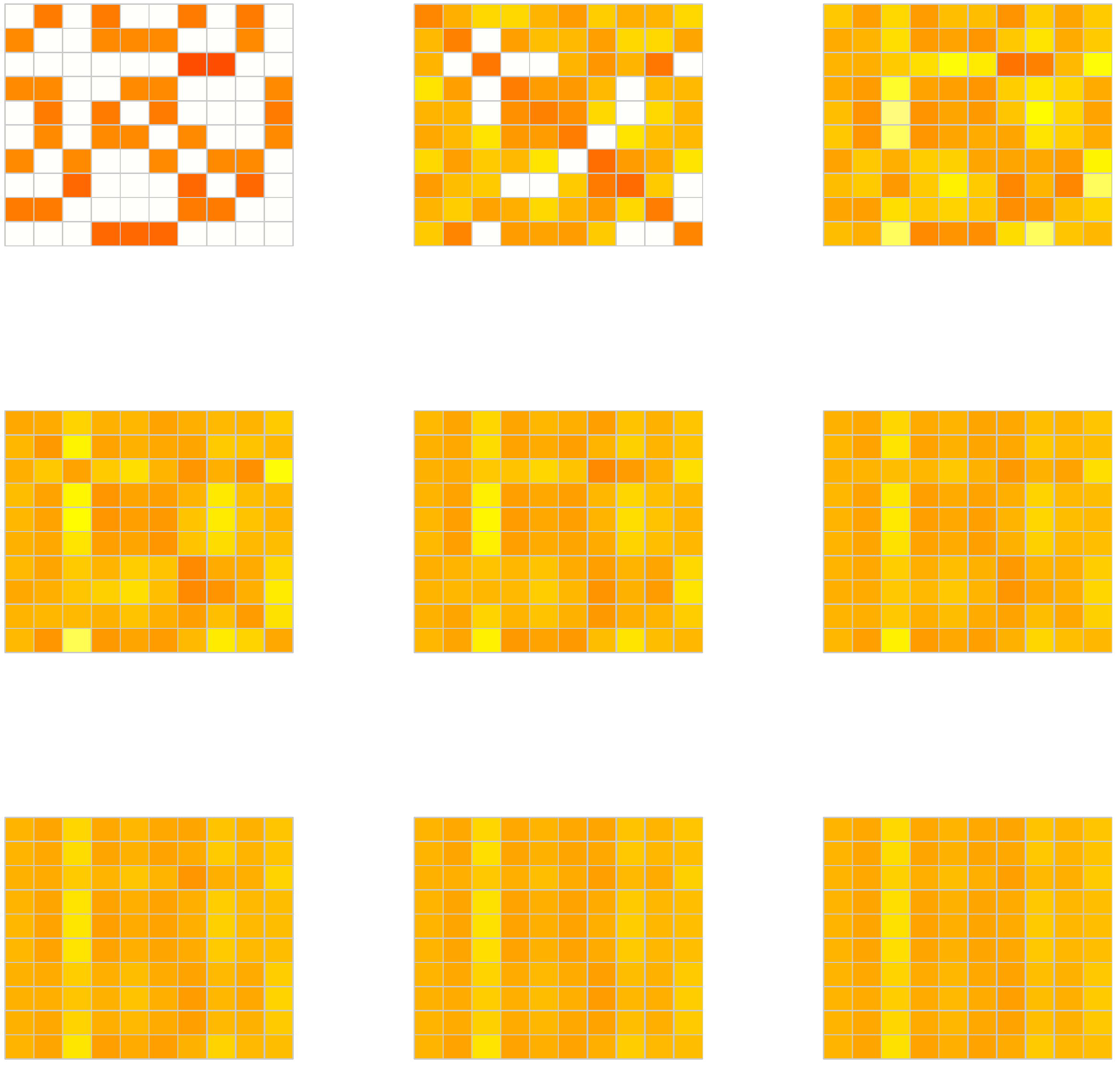}
\end{center}
\caption[Draw-Wise Sampling Probabilities for With-Replacement Process]{Draw-wise sampling probabilities for with-replacement random walk process on network, conditional on starting node, for steps 1 through 9.} \label{withexclude0228b}
\end{figure}

To illustrate the convergence to this distribution, consider the $n$-step transition probabilities for a random walk on the simple example network depicted in Figure \ref{sociogram0228b}.  If we use the color mapping of probabilities depicted in Figure \ref{cdcheatprobs}, we can represent these probabilities as colored matrices in which the color of the $\{i,j\}^{th}$ cell in the $n^{th}$ transition matrix represents the probability that a random walk beginning at node $i$ lands on node $j$ at the $n^{th}$ step.  For the first nine steps, $n=1, \ldots 9,$ of a random walk on the graph in Figure \ref{sociogram0228b} these matrices are given in Figure \ref{withexclude0228b}.

The non-solid column colors early in the process represent dependence of the sample on the seed values.  The more solid columns in later steps represent the convergence of the process to draw-wise sampling probabilities independent of starting node.  That is, in step 4, for example, the third row contains the darkest cell in the ninth column.  This indicates that the fourth step of a random walk beginning at node 3 is more likely to be node 9 than the fourth step of a similar walk beginning at any other node.  By the ninth step, however, the nearly solid-colored ninth column indicates near independence of the starting node.  That is, whichever node the process started with, by the ninth step, the ninth node has about the same probability of selection.
The resulting draw-wise sampling probabilities converge to probabilities proportional to nodal degree, so that by the last plot of Figure \ref{withexclude0228b}, the solid colors approached by the columns represent probabilities proportional to the degrees of the corresponding nodes.
In a larger network with a more complex structure, this type of convergence is likely to require a great many more steps, so that it is unreasonable to expect the convergence of nodal sampling probabilities in the population sizes and number of waves typically occurring in RDS.

Based on this random walk model, \cite{volzheck08} introduce a modified Hansen-Hurwitz \citep{hh43} estimator for $\mu$:
\bea
\hat{\mu}_{\rm VH} = \frac{\sum_{i: S_i=1}\frac{z_i}{\tilde{d_i}}}{\sum_{i: S_i=1}\frac{1}{\tilde{d_i}}},
\label{volzheckest}
\eea
where the self-reported nodal degree $\tilde{d_i}$ is used to approximate the true nodal degree $d_i$. The form of this estimator is identical to that of the generalized Horvitz-Thompson estimator in (\ref{rdsht}).  Due to the ratio format of this estimator, the sampling probabilities are required only up to a constant of proportionality.  The estimator should take the form:
\bea
\hat{\mu} = \frac{\sum_{i: S_i=1}\frac{z_i}{p_i}}{\sum_{i: S_i=1}\frac{1}{p_i}},
\label{hhpis}
\eea
where $p_i$ is the draw-wise selection probability of unit $i$.  In the case where $p_i \propto \tilde{d_i}$, (\ref{hhpis}) reduces to (\ref{volzheckest}).

\subsection{Assumptions of the Current Estimator}
For the validity of the Volz-Heckathorn (V-H) estimator, it is sufficient that the nodal sampling probability $\pi_i$ of each node $i$ is proportional to that node's self-reported {\it degree}, $\tilde{d_i}$, or number of {\it alters} in the target population.

These sampling weights are derived from the above model of a single non-branching, with replacement, random walk process at equilibrium.  This model is known to be a simplification; the process is branching, without replacement, and does not begin at, or even converge to a fixed equilibrium.   In addition to these known approximations, estimation requires a set of network and sampling assumptions to remove the bias induced by seed selection even in the ideal random walk setting.  In addition to the requirements of theoretical convergence, a final set of assumptions regarding respondent behavior is required to allow for the estimation of inclusion probabilities as proportional to reported nodal degrees.  Each set of assumptions is discussed in turn, and the full array of assumptions is indicated in Table \ref{tab:assred}.

\begin{table}\caption[Assumptions of the Volz-Heckathorn Estimator]{Assumptions of the Volz-Heckathorn Estimator, (assumptions in blue are considered in the simulation study in Section \ref{section:simstudy})}
\begin{center}
\begin{tabular}{l||c|c}
& Network Structure  & Sampling Assumptions\\
& Assumptions & \\
\hline
\hline
Random Walk & \ecol \eone Network size large ($N >> n$) & \ecol \eone With-replacement sampling \\
Model & 
& \eone Single non-branching chain \\
\hline
Remove Seed & \ecol \etwo Homophily weak enough & \ecol \etwo Sufficiently many sample waves \\
Dependence & Connected graph & \\
\hline
Respondent   & All ties reciprocated & Degree accurately measured  \\
Behavior & & \ecol \etwo Random referral   \\
\end{tabular} \label{tab:assred}
\end{center}
\end{table}

The distinction between the Hansen-Hurwitz estimator in (\ref{volzheckest}) and the Horvitz-Thompson estimator in (\ref{rdsht}) is the first indication of a deviation from the assumptions of the model.  Although the forms of the Horvitz-Thompson and Hansen-Hurwitz estimators are similar, they differ in that the first is based on the overall or {\it list-wise} inclusion probabilities, $\pi$, of a typically without-replacement process.  That is, the list-wise inclusion probability, $\pi_i$ of node $i$ is the probability that $i$ is sampled at any point.   The Hansen-Hurwitz estimator, however is based on the {\it draw-wise} selection probabilities, $p$, of a {\it with-replacement} process, such that $p_i$ is the probability that node $i$ is sampled on any given selection, assumed constant across selections.  In respondent-driven sampling, as with most social science sampling procedures, population members are not sampled more than once.  The process is without replacement.  To finesse this difference, Volz and Heckathorn require the population or {\it network size to be large} with respect to the sample size, (i.e. a small sample fraction).

In seeming opposition to the notion of a small sample fraction is the need for {\it sufficiently many waves of sampling} to remove the bias induced by the convenience sample of seeds.  Note that the stationary distribution of the Markov Chain occurs only after convergence is attained, which will occur only after many waves from the initial sample.  This process will be facilitated by a highly-connected, non-clustered network.  In particular, to reduce the bias of RDS estimators, the network structure should not be highly clustered on $z$, or equivalently have only  {\it weak homophily} structure on $z$. A {\it connected graph} is also required for an irreducible Markov Chain.   Note that convergence also requires an aperiodic transition structure.  As noted in \cite{salgheck04}, this condition is satisfied for any connected graph with at least one triangle, as is highly likely in an RDS setting.
Of course, the notion of convergence is fundamentally problematic for a without-replacement process.

Even if we assume a process well approximated by a converged Markov chain, we must make assumptions about the behavior of respondents in order to estimate the selection probabilities necessary for inference.  To begin with, we require the assumption that respondents {\it accurately report their degrees} (i.e. $\tilde{d_i}=d_i$).  We also require $p_i \propto \tilde{d_i}$.  For this, we need a random walk on a network in which {\it all ties (relations) are reciprocated} (i.e. an undirected graph).  If relations in the network are not reciprocated, selection probabilities are determined by a more complex function of the graph structure, which depends on unobserved features of the relations in the population.
Finally, we require the {\it random distribution of coupons among alters}.  Where coupons are distributed other than at random, selection probabilities will be dependent on the characteristics determining the distribution, and are likely either unobservable or difficult to include in an estimator.

\subsection{The With-Replacement Assumption}

Because the without-replacement nature of RDS sampling is critical to performance and is poorly understood, we consider this aspect in greater detail.
A salient difference between with-replacement and without-replacement sampling is the different emphases on draw-wise and list-wise sampling probabilities.  In a with-replacement random walk on a graph, the marginal probability of the $k^{th}$ sample being node $j$ is constant after the process has converged.  In this case, it is also possible to consider the probability that node $i$ is sampled during the course of a chain of length $k$, or to consider the ratio of expected frequencies of nodes $i$ and $j$ in a sample of size $k$.

Now consider the case of a self-avoiding (i.e without-replacement) random walk on a graph.  Then it is natural to consider the probability of including node $j$ in a path of length $k$ originating at node $i$.  Because this process is without-replacement, there is no equilibrium distribution of draw-wise sampling probabilities.  It is therefore less natural to think about the probability of sampling node $j$ at the $k^{th}$ step of a path originating at node $i$.

In the current RDS estimator, the estimated weights are based on a draw-wise approximation in a with-replacement sampling model, while the true selection probabilities are based on a without-replacement process, for which draw-by-draw weights are not germane.

Furthermore, the sampling weights are based on the notion of the process occurring at equilibrium.  A without-replacement process, however, does not have equilibrium sampling weights.  Even if the process were with-replacement, the process does not begin at equilibrium.
Current RDS analysis does not discard more than perhaps the seeds in the final analysis, while a standard stochastic process analysis typically discards a large number of initial waves before selecting the probability sample to be analyzed.
Also, there is no evidence that the number of waves sampled in RDS samples is reasonable for approaching an equilibrium distribution on nodes.

These concerns highlight the importance of considering the without-replacement structure of RDS sampling.  This feature is a key advantage of our simulation approach.

\subsection{Estimators of Uncertainty}
Uncertainty estimates in the current RDS software \citep{rdsat} are provided by a bootstrap procedure proposed by \cite{salg06}.  In this procedure, multiple bootstrapped replicates of the sample are created by re-sampling nodes.  The procedure used maintains some faithfulness to the dependence in the sample by
sampling subsequent nodes from the subset of nodes referred by others with the same $z_i$ value as the current node.  Seeds in the replicates are chosen completely at random from among the sampled nodes. The estimator is computed on each replicate sample, and a 90\% confidence interval is then based on the percentiles of the re-sampling distribution.  This estimator makes a first order correction for the homophily structure, but does not address other features of the sampling process.
In the case of strong homophily and biased seeds, for example, the entire sample, and consequently most replicates, could drastically over-represent the nodes most similar to the seeds, resulting in poor coverage rates of the resulting confidence interval.

\cite{volzheck08} present an alternative analytical variance estimator.  Their approach involves treating the sample as successive draws from a Markov process, this time on the state space of discrete classes of $z_i$.  Based on a simulation study, they conclude that their estimator is slightly conservative, however the study is conducted in a with-replacement frame with $N=10,000$ and $n=500$.
It is unclear how this estimator would perform under the considerably more complex conditions closer to the true RDS sampling setting.

\section{Simulation Study to Evaluate Sensitivity to Assumptions} \label{section:simstudy}

Our simulation study is designed to isolate the effects of several features of RDS study design and implementation.  To make our study as realistic as possible, we chose most parameters to match the characteristics of the pilot data from the CDC surveillance program \citep{aqcdc06} as nearly as possible.  The general procedure was as follows:
\begin{enumerate}
 \item Simulate 1000 networks with a specified structure
 \item Implement a variant of the RDS sampling process with desired characteristics on each network
 \item Compute the V-H estimator of the population proportion from each sample
 \item Compare each estimate to the known true population proportion
 \item Repeat steps 1-4 with variants of the network structure and sampling process.
\end{enumerate}

In particular, simulation studies were conducted to evaluate the sensitivity to assumptions listed in blue in Table \ref{tab:assred}.  These correspond to two factors important to removing seed bias: sufficiently many waves of sampling, and homophily weak enough, as well as an evaluation of the strategy of treating seed bias by basing estimators on later waves only.  We examine deviations from ideal respondent behavior by considering non-random referral.  Then we consider deviations from the with-replacement random walk model in terms of the assumptions of network size large (compared to sample), and with-replacement sampling. Each assumption is examined by comparing the estimators resulting from at least two variants on a basic simulation study design.  We present the general design first, then introduce specific variants with each specific sub-study.

Because the CDC's surveillance system aims for a sample size of 500, and many RDS studies approach exhaustion of their populations of interest, we consider a baseline population size of 1000 for this series of studies.  All simulations considered samples of size 500.  We also consider a mean degree of 7, close to the mean of the pilot data from the CDC study \citep{aqcdc06}.

The primary aim of RDS analysis is the estimation of population proportions in a hidden population.  Therefore, we assign a discoverable class to each member of the simulated population.  In reference to studies designed to estimate the prevalence of infectious disease, we refer to this characteristic as {\ql}infection status.\qr  We assign the {\ql}infected\qr status to 20\% of simulated population members.

For the primary step of the procedure we need to be able to generate
networked populations with a pre-specified population size and controllable
homophily, relative activity levels, and mean degree. We do this by
representing the structure with an Exponential family Random Graph Model
(ERGM)\citep{sprh06}. Here the relations $y$ are represented as a realization
of the random variable $Y$ with distribution:
\begin{eqnarray}
P_{\eta}(Y=y | x) = \exp\{\eta{\cdot}g(y,x)-\kappa(\eta,x)\}\quad \quad y\in {\cal
Y},
\label{ergm}
\end{eqnarray}
where $x$ are dyadic-level covariates, $g(y,x)$ is a $p$-vector of network statistics,
$\eta\in \IR^p$
is the parameter vector, ${\cal Y}$ is the
set of all possible undirected graphs, and
$\exp\{\kappa(\eta,x)\} =
\sum_{u\in{\cal Y}}\exp\{\eta{\cdot}g(u,x)\}
$
is the normalizing constant \citep{bar78}.
The structure of the networks represented is determined by the choice of
$g(y,x)$.

A key complicating feature for RDS analysis is the tendency for homophily in the formation of relations, and most acutely when the homophily is on the characteristic of interest.  For this reason, we induced homophily on infection status in these simulations.  To promote the interpretability of the network features, we
 control homophily by controlling the relative probability of an edge between two infected nodes, and the probability of an edge between an infected and an uninfected node.  For most of the simulations,  the edge probability between the two infected nodes was fixed at five times that of the mixed dyad.  In the interest of keeping the mean degree constant across the two groups, this, along with the 20\% infected,  implies that an edge between two uninfected nodes is twice as likely as an edge in a mixed dyad.

For some of the sub-studies, it was important to control the relative activity level of infected and uninfected population members.  For these studies, we varied the relative activity level of the two groups, keeping the relative probabilities of edges in infected-infected and uninfected-infected dyads constant.  Unless otherwise noted, the activity levels of infected and uninfected nodes were the same.

These features were represented in the ERGM by choosing
network statistics to represent the mean degree, the relative
activity levels of the two groups, and homophily (based on
using the {\ql}infected{\qr} status as a dyadic covariate).
The parameter $\eta$ was chosen
so the expected values of the statistics were equal to the values given
above \citep{vanduijngilehan09}.
This was implemented in {\tt statnet} \citep{statnet}.

\begin{figure}[h]
\begin{center}
    \includegraphics[width=4.5in,height=4.5in]{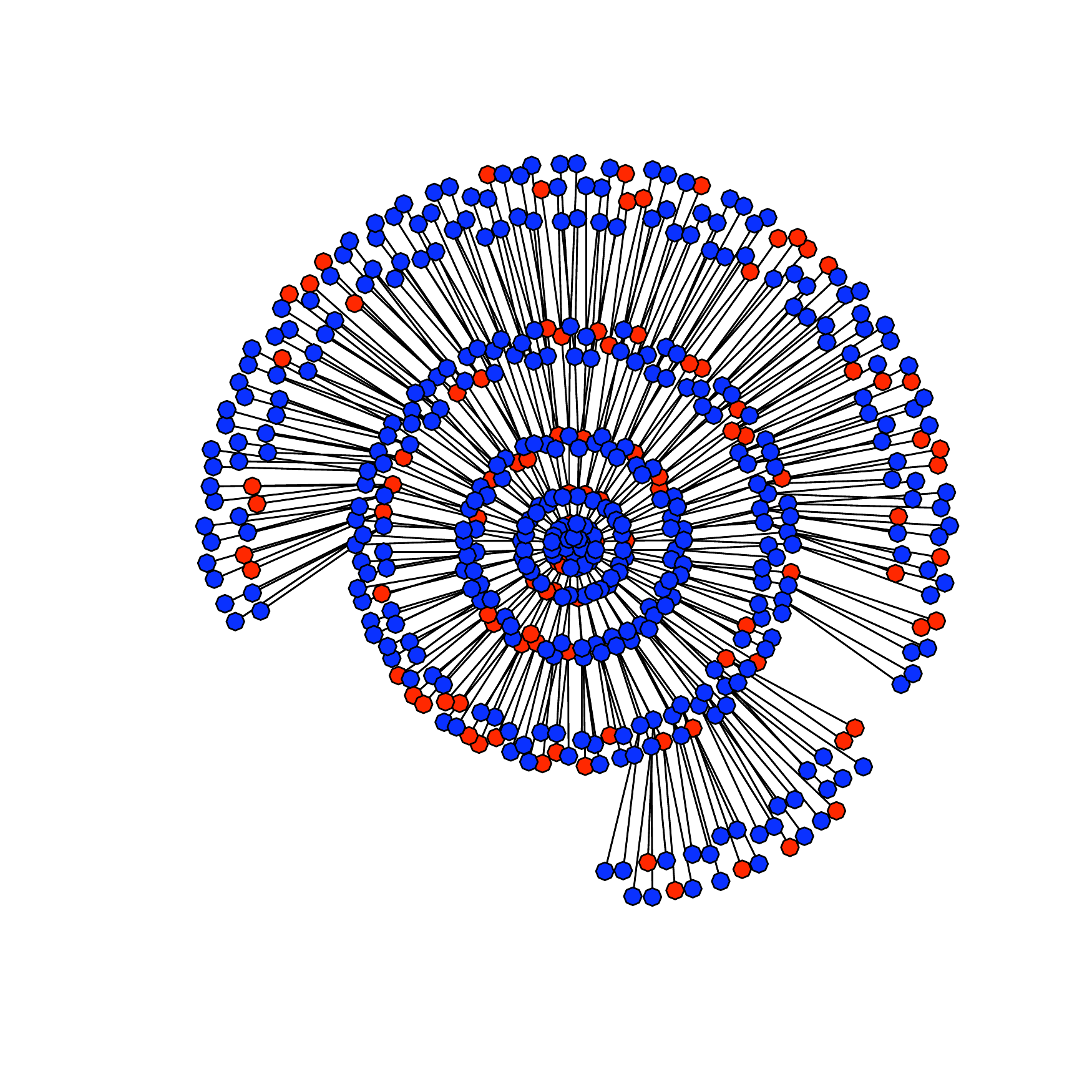}
\end{center}
\caption{Illustration of simulated sample beginning with 10 seeds (inner circle), including 2 infected (red), and continuing through four full and a fifth partial wave to obtain a sample of 500.} \label{samplepicture}
\end{figure}

The RDS sampling mechanism is again designed to mimic that of the CDC's pilot study.   A baseline of 10 seed nodes were chosen for each sample.  To make the strongest case for RDS, the seed nodes were chosen with probability proportional to degree\footnote{The true selection process is sequential probability proportional to degree, where each successive node is selected {\it from the previously un-sampled nodes} with probability proportional to degree.  For samples of size 10 or 20 out of 1000 nodes, the difference between these two procedures is negligible.}.  Because a primary goal of RDS is to eliminate seed bias, seeds were chosen in three ways:  completely at random with respect to infection status, completely from within the infected sub-population, or completely from within the uninfected sub-population.  Subsequent sample waves were selected without-replacement by sampling up to two nodes at random from among the un-sampled alters of each sampled node.  Exactly two alters were sampled whenever two or more un-sampled alters were available.  This process typically resulted in the sampling of four complete waves and part of a fifth wave, stopping when a sample size of 500 was attained. One sample obtained with this standard set of parameters is illustrated in Figure \ref{samplepicture}.

In the following sections, we discuss simulations to address the assumptions in blue in Table \ref{tab:assred}. The sub-sections are named after the assumptions they address.  Note that for comparison, we keep the vertical range fixed across all comparable box-plots. Therefore, data in some plots are not fully visible in this standard plot range.  For completeness, any truncated plots are displayed in replicated plots in the Appendix,
with ranges expanded to include all data.

\subsection{Removing Seed Bias}
We consider two features which determine the effect of seed selection on the bias of RDS estimators:  the number of waves of sampling and the degree of homophily in the underlying population of relations.
If mixing were completely random, the first wave of sampling after the seeds would already be independent of the seed values.  On the other hand, for any level of homophily in which the network still consists of a {\it single connected component}, a full-wave sample (i.e., continuing sampling until exhaustion) would be independent of seed bias.  In practice, there is at least some homophily in the population, and the sample never captures the full population, so the combination of these two features determine the amount of seed bias.

In this section, we also include an evaluation of the strategy of basing the estimator on later waves of the sample only in order to remove those samples most dependent on the convenience sample of seeds.

\subsubsection{Sufficiently Many Waves of Sampling} \label{section:waves}

Section \ref{sec:estimator} described the discrepancy between the number of waves of sampling necessary for convergence of a random walk process and the number of waves typically sampled in a RDS study.  Clearly, the Markov Chain equilibrium-based assumptions about the inclusion probabilities of individual nodes are not supported.  Given the complexities of the actual sampling process, it is of interest to understand the effect on estimator performance of more or fewer sampling waves.  This is practically important because RDS researchers need to understand the relative importance of sampling designs that encourage many waves of sampling.

To evaluate the difference in performance with {\it more} or {\it fewer waves} of sampling, we consider two scenarios:  6 seeds, contributing to 500 samples from 6 waves ({\it many waves}, $k=6$), and 20 seeds contributing to 500 samples from 4 waves ({\it few waves}, $k=4$).  Furthermore, we consider each of these regimes under three scenarios of seed selection:  {\it all uninfected}, {\it random with respect to infection}, and {\it all infected}.

\begin{figure}[h]
\begin{center}
    \includegraphics[width=4.5in,height=4.5in]{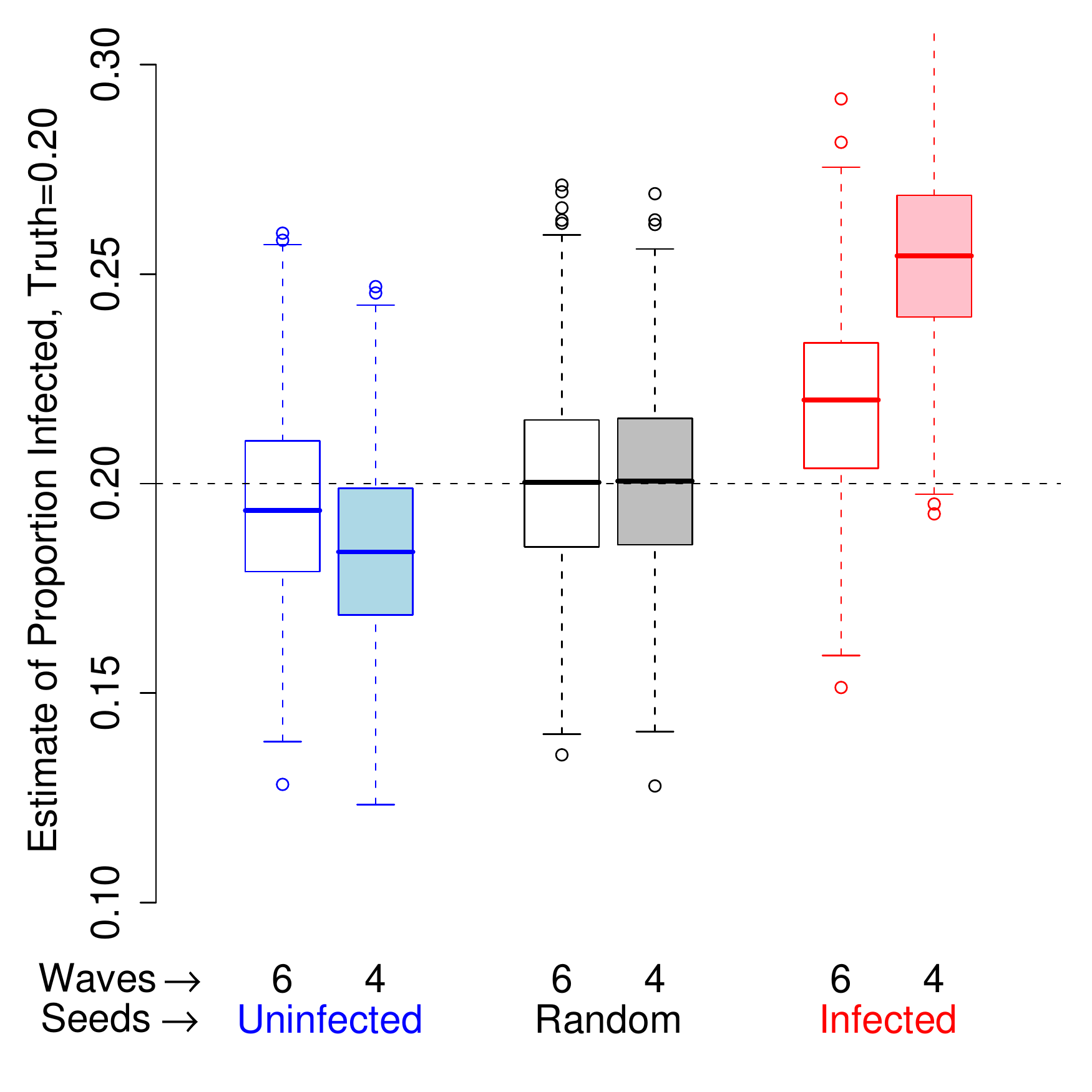}
\end{center}
\caption[Comparison of V-H Estimators with Many and Fewer Sampling Waves]{V-H estimators from samples of 6 seeds, 6 waves (first, third, and fifth boxes), and 20 seeds, 4 waves, from seeds selected from all uninfected nodes (first two boxes), random nodes with respect to infection (second two boxes), and all infected nodes (last two boxes).} \label{cdcwave}
\end{figure}

The distributions of the V-H estimators in each case are summarized in the box-plots in Figure \ref{cdcwave}.  The first thing to note in this figure is that the middle two boxes are very similar.  Neither shows any appreciable bias, and the variances are very similar.
This is because when the seeds are chosen at random with respect to infection status, this constitutes sampling from close to the theoretical equilibrium distribution of the process.  

The strong performance of the {\it fewer waves} sample does not carry over to the cases with biased seed selection.  For both the infected and uninfected seed simulations, the bias induced by seed selection in the {\it fewer waves} simulation is substantially lower than that in the {\it more waves} simulation.  This is because shorter sampling chains lead to increased overall dependence of the sample on the seeds, and therefore lead to greater bias in the estimators whenever the seeds are biased.

Note that there is more bias in the case of the {\it all infected} seeds than for case of {\it all uninfected} seeds.  This is because the infected nodes form a smaller group with equal activity level to the uninfected nodes.  Therefore, infected nodes have infected alters at a rate 5 times that indicated by random mixing, while uninfected nodes have uninfected alters at a rate only twice as often as indicated by random mixing.  Therefore, if referral is random, infected nodes recruit disproportionately from within-group, contributing to a slower rate of transition from infected to uninfected than from uninfected to infected.

\subsubsection{Homophily Weak Enough}\label{section:homoph}

To evaluate the difference in performance with greater and lesser homophily, or clustering by infection status, we compare estimators for two different levels of homophily.  The standard level of homophily, described in Section \ref{section:simstudy} is treated as the {\it lower level}.  For the {\it higher level}, we consider the case where an edge between two uninfected nodes is four times as likely as an edge between an infected and uninfected node.  To maintain the same mean degree for both subgroups, this implies that the probability of an edge between two infected nodes is 13 times that of an infection-discordant edge.
We again consider each of these regimes under the three seed selection scenarios:  {\it all uninfected}, {\it random with respect to infection}, and {\it all infected}.

\begin{figure}
\begin{center}
    \includegraphics[width=4.5in,height=4.5in]{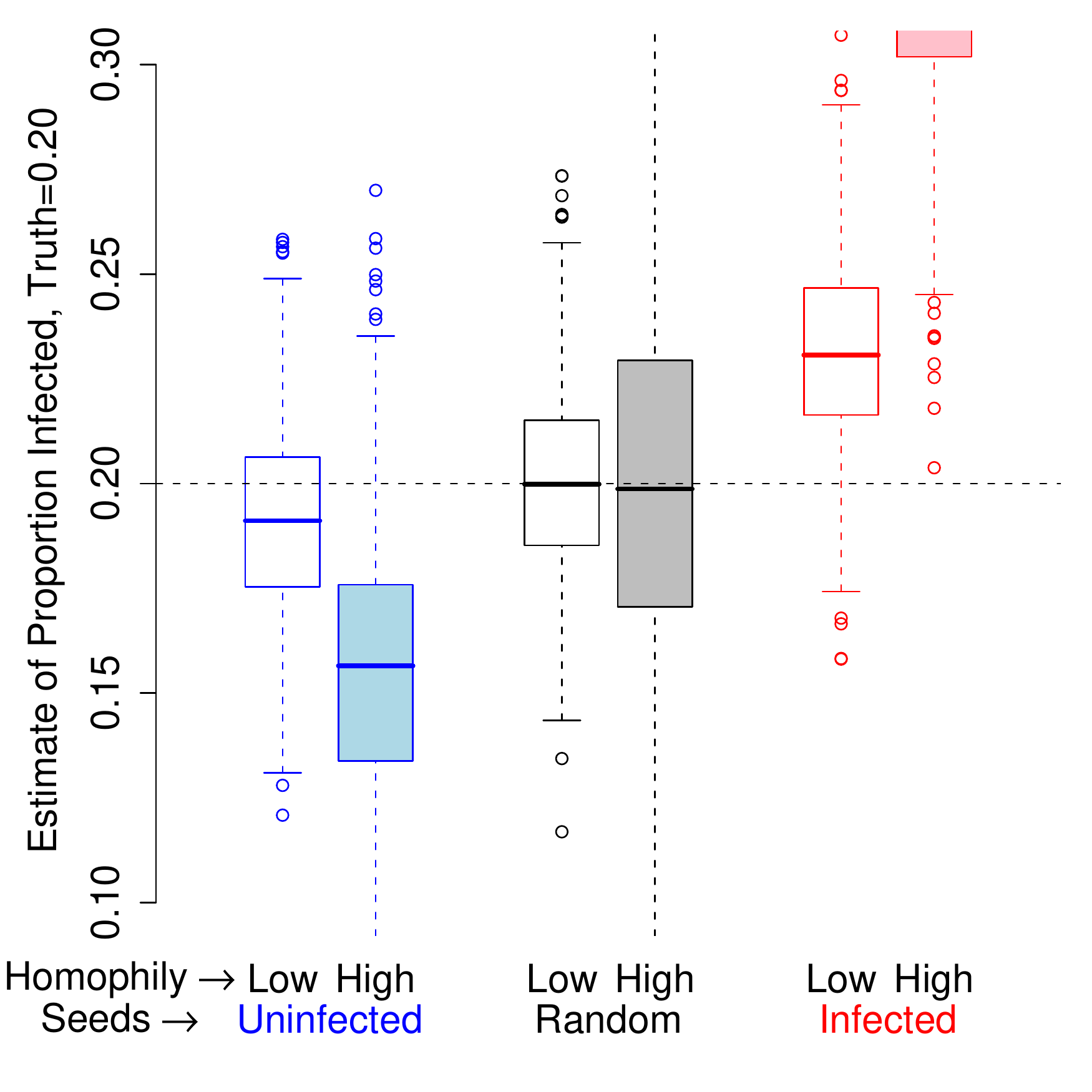}
\end{center}
\caption[Comparison of V-H Estimators with Standard and Elevated Homophily]{V-H estimators from samples with standard (first, third, and fifth boxes), and elevated homophily, from seeds selected from all uninfected nodes (first two boxes), random nodes with respect to infection (second two boxes), and all infected nodes (last two boxes).} \label{cdchomoph}
\end{figure}

Figure \ref{cdchomoph} summarizes the results of these simulations.  The case of unbiased seeds  illustrates the relationship between homophily and variance of the estimator.
The higher homophily condition shows 3.5 times the variance (1.9 times the standard deviation)
of the lower homophily condition.  This is because in the high homophily condition, the higher correlation between the infection statuses of successively sampled nodes leads to less information for comparable sample sizes.

In the biased seed conditions, there is both a bias and a variance difference between the two homophily levels.  The higher homophily conditions show far greater bias than their biased-seed but lower homophily counterparts.  Again, this is to be expected, as the higher homophily induces higher dependence between the characteristics of the seeds and the characteristics of the subsequent samples.

\begin{figure}[h]
\begin{center}
    \includegraphics[width=2in,height=2in]{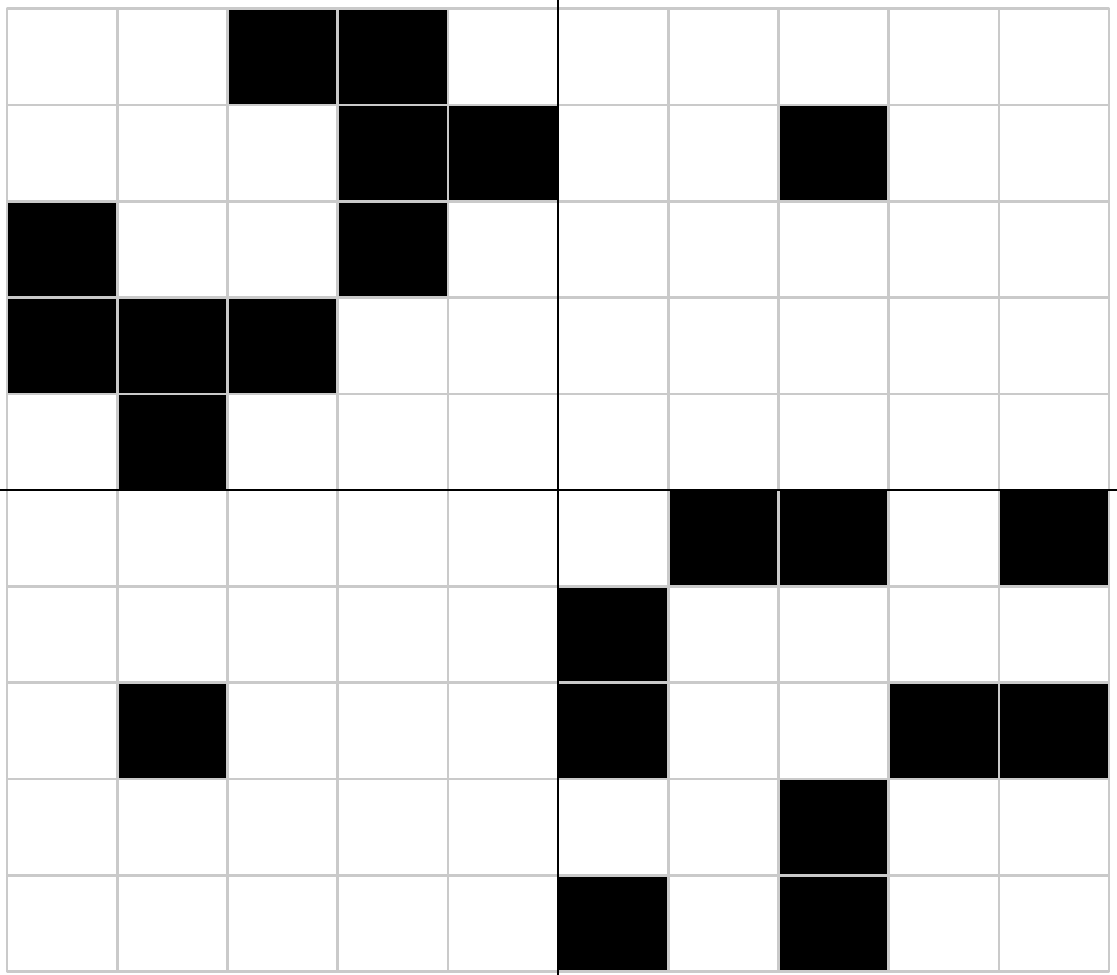}
\end{center}
\caption[Sociomatrix for Network of Size 10 With High Homophily]{Schematic depiction of highly clustered sociomatrix for network of size 10.} \label{cdcMCHL}
\end{figure}

Both the bias and the variance increases in the higher homophily condition can be understood in terms of the mixing process.  We can visualize the slower mixing of a link-tracing process on a high-homophily network with a visualization similar to that in Figure \ref{withexclude0228b}.  Consider the high-homophily network of 10 nodes whose sociomatrix is depicted in Figure \ref{cdcMCHL}.\knote{haven't introduced notion of sociomatrix - should use sociogram here, or introduce sociomatrix}
\mnote{sociomatrix is defined implicitly on page 6. I do not think more is
needed here.}

Using the color mapping of probabilities depicted in Figure \ref{cdcheatprobs}, we again represent the 1, 5, 10, and 14-step transition matrices, this time in Figure \ref{cdcsteps51014HL}.  By step 14, the columns appear nearly homogeneous in color, there is a clear block diagonal pattern to these matrices, which persists in step 14.  This pattern illustrates that a sample starting with one of the first five nodes (the first cluster) is more likely to land on another node from that cluster in the first, fifth, tenth, or fourteenth step.   The stronger the clustering pattern, the more persistent the bias induced by seed selection in subsequent samples.

\newcommand{\apples}{1.9}
\begin{figure}
\begin{center}
\subfigure[1-step]
{
    \includegraphics[width=\apples in,height=\apples in]{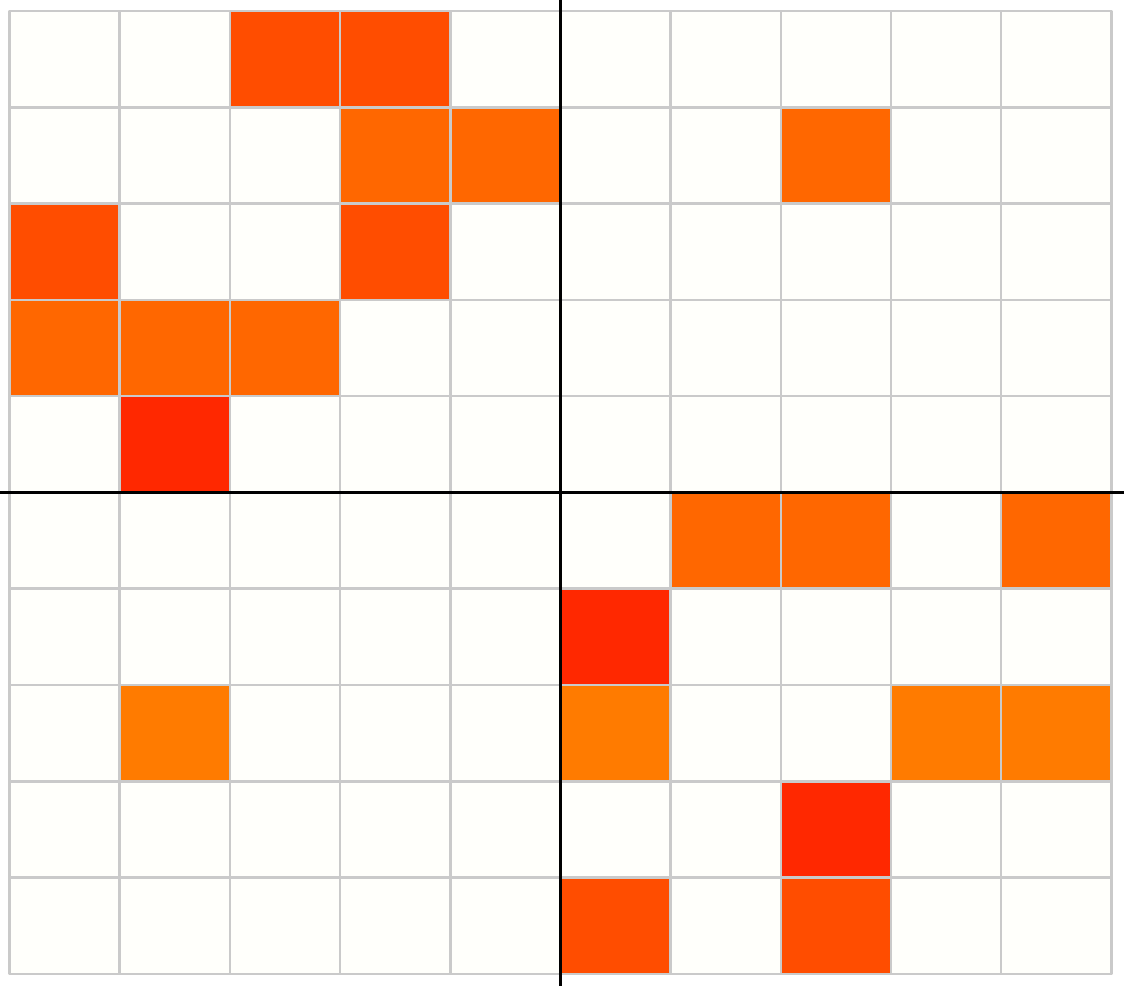}
} \hspace{.25cm}\subfigure[5-step]
{
    \includegraphics[width=\apples in,height=\apples in]{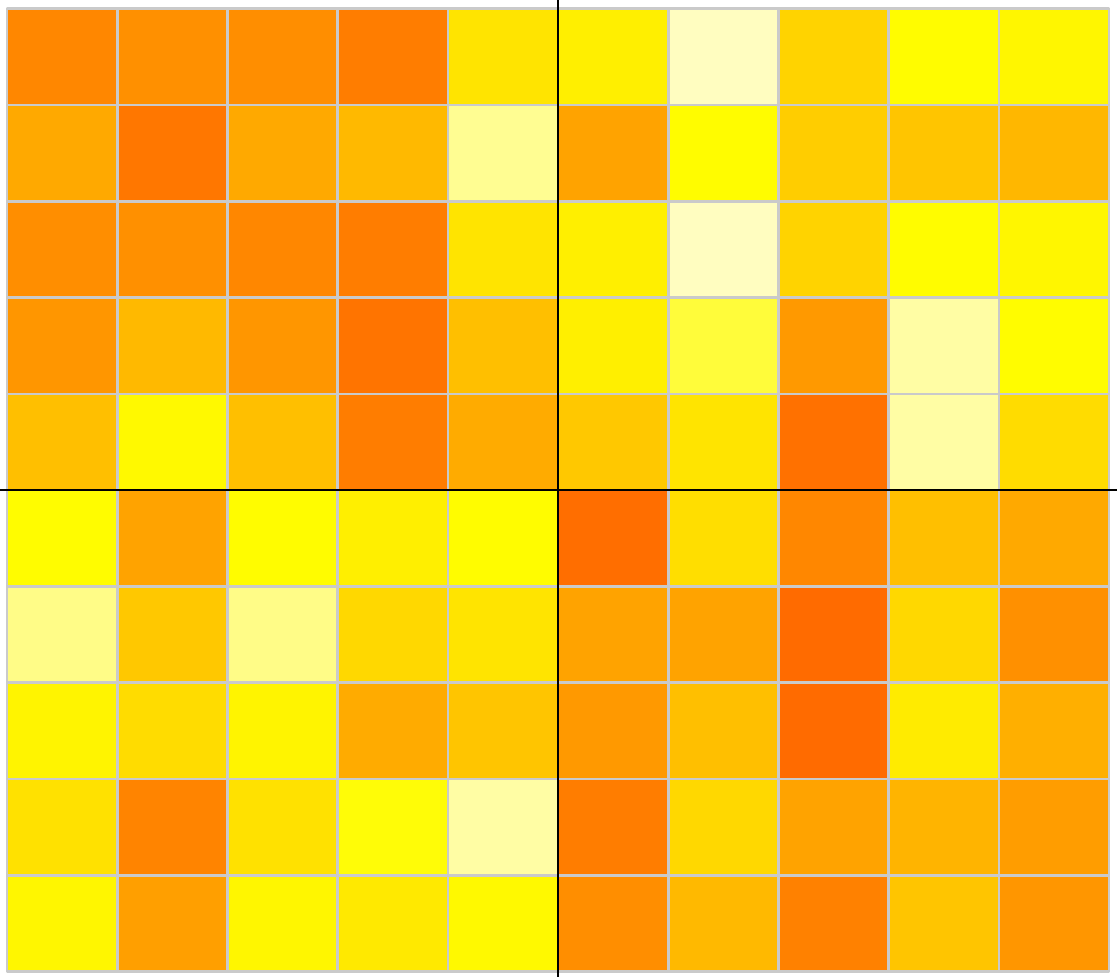}
} \hspace{.25cm}
\subfigure[10-step]
{
    \includegraphics[width=\apples in,height=\apples in]{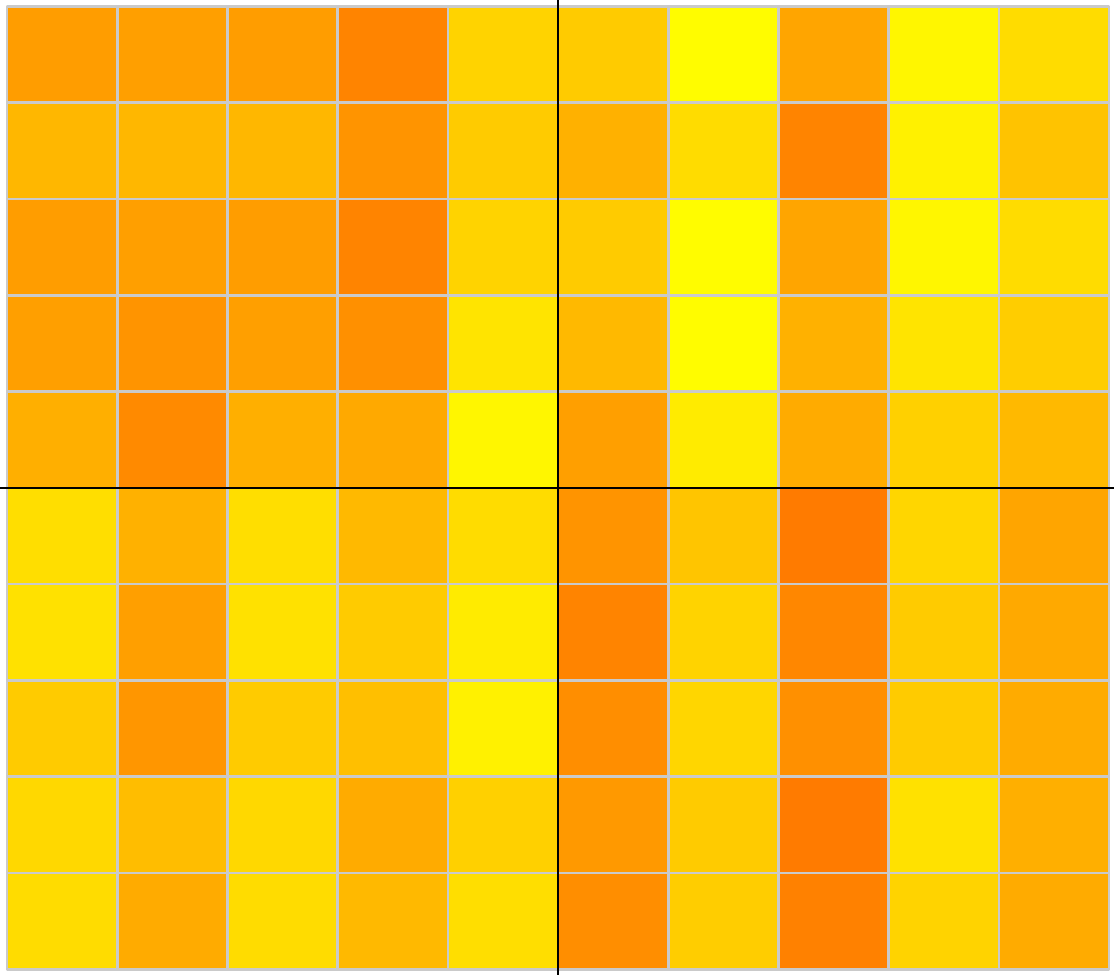}
} \hspace{.25cm}
\subfigure[14-step]
{
    \includegraphics[width=\apples in,height=\apples in]{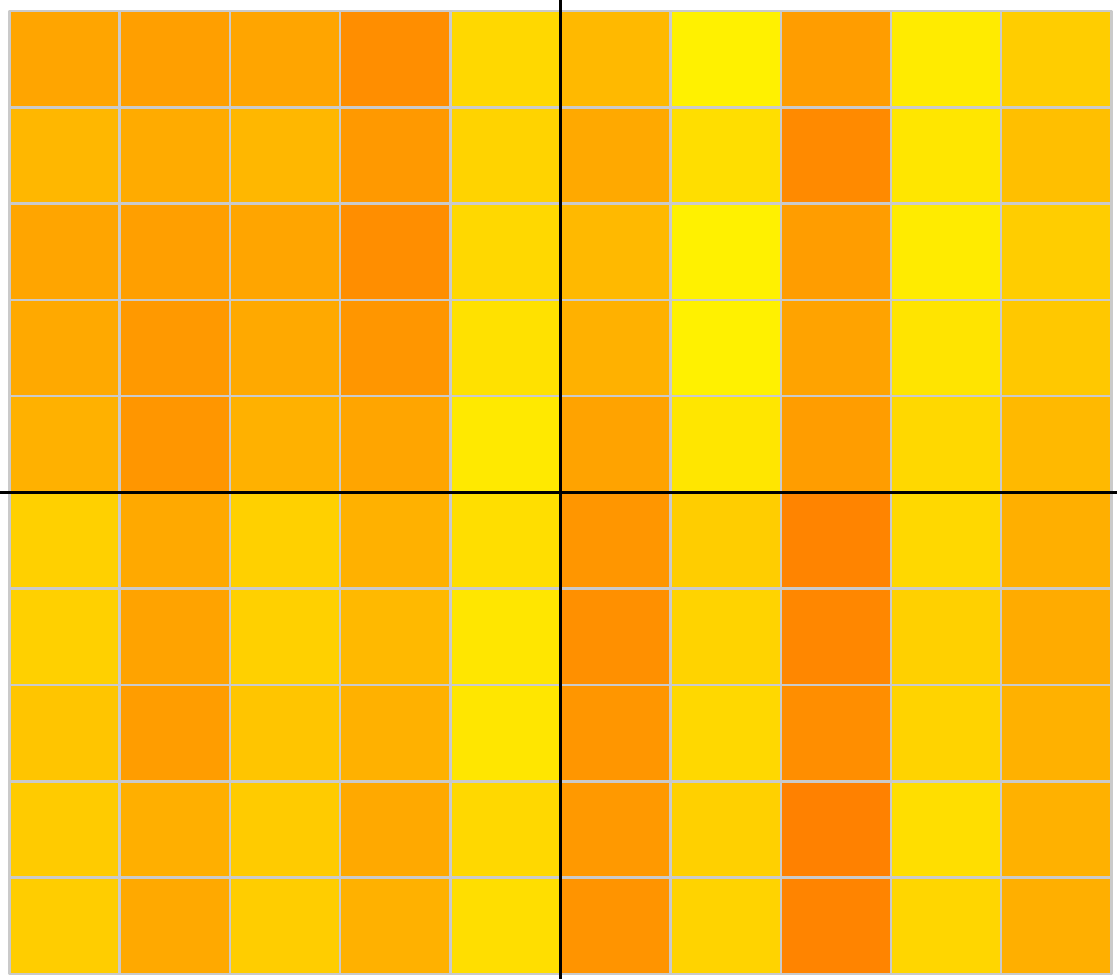}
}
\end{center}
\caption[Five, Ten, and Fourteen Step Transition Matrices for Network With High Homophily]{One, five, ten, and fourteen step transition probabilities for highly clustered network.} \label{cdcsteps51014HL}
\end{figure}

\subsubsection{Estimation based on later waves only} \label{section:earlier}

One strategy for reducing the seed dependence in RDS is to base estimation on only nodes sampled after a specified wave.
In this section, we demonstrate that bias
may not be dispelled from RDS samples by discarding early waves,
and that in a without-replacement setting, this approach can even introduce bias.

Under the random walk model, it is known that earlier waves are more dependent on seed bias than later waves.  Typically with Markov Chains, earlier waves are discarded, and analysis conducted only on later samples.  In RDS, however, analysis is typically conducted on all data collected, possibly discarding the seeds.  This juxtaposition suggests that RDS estimation might be improved by discarding earlier waves of the sample.

We begin with the study comparing the performance of estimators with few and many waves of sampling described in Section \ref{section:waves}, and illustrated in Figure \ref{cdcwave}.  We then consider the performance of the estimator when only the seeds, the seeds and first wave, the seeds, first and second waves, and the seeds and first three waves are discarded, and analysis is conducted on the remainder of the sample.

\begin{figure}[h]
\begin{center}
\subfigure[Discard Seeds]
{
    \label{withouts}
    \includegraphics[width=\sss,height=\ttt]{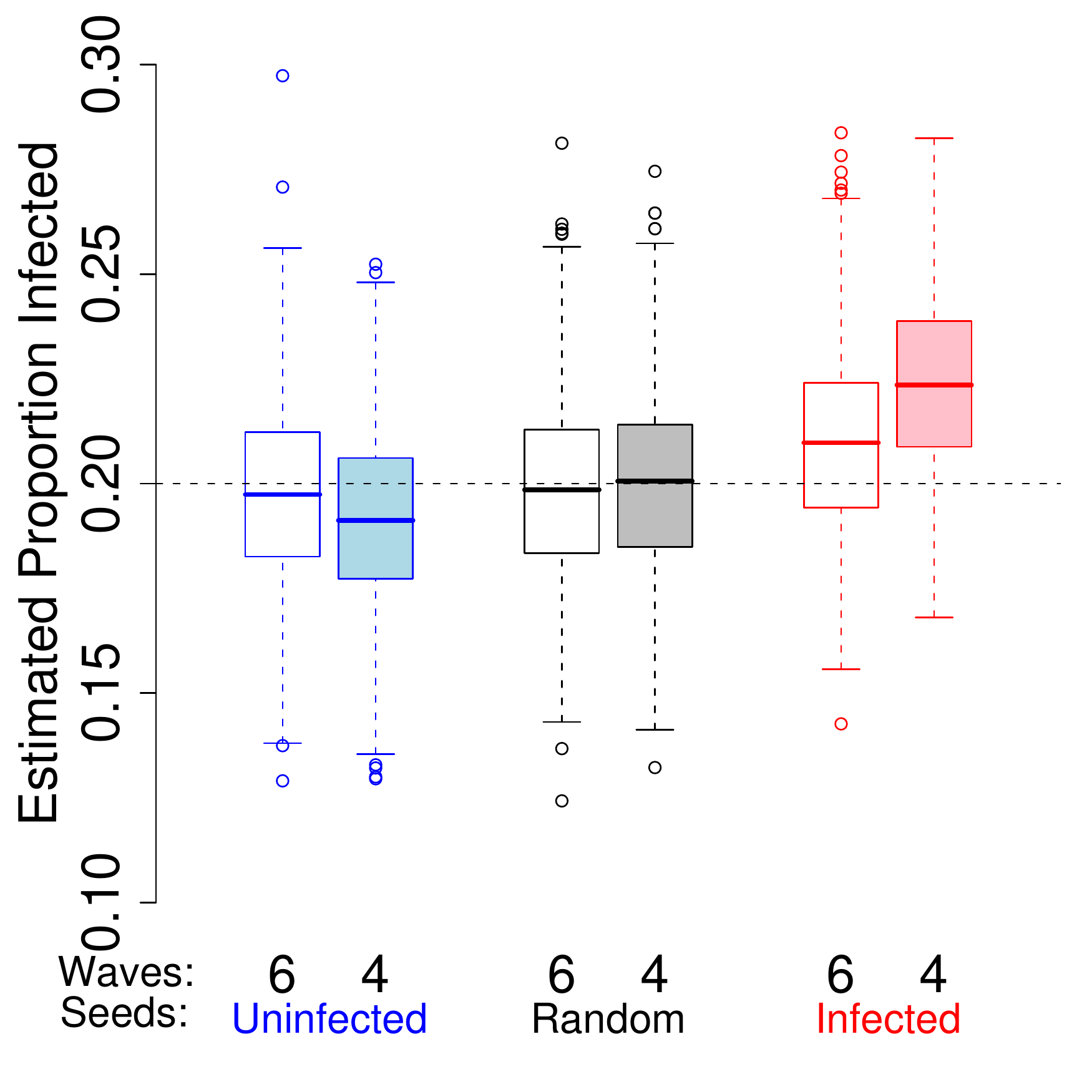}
} \hspace{.5cm}
\subfigure[Discard Seeds, Wave 1]
{
    \label{withouts1}
    \includegraphics[width=\sss,height=\ttt]{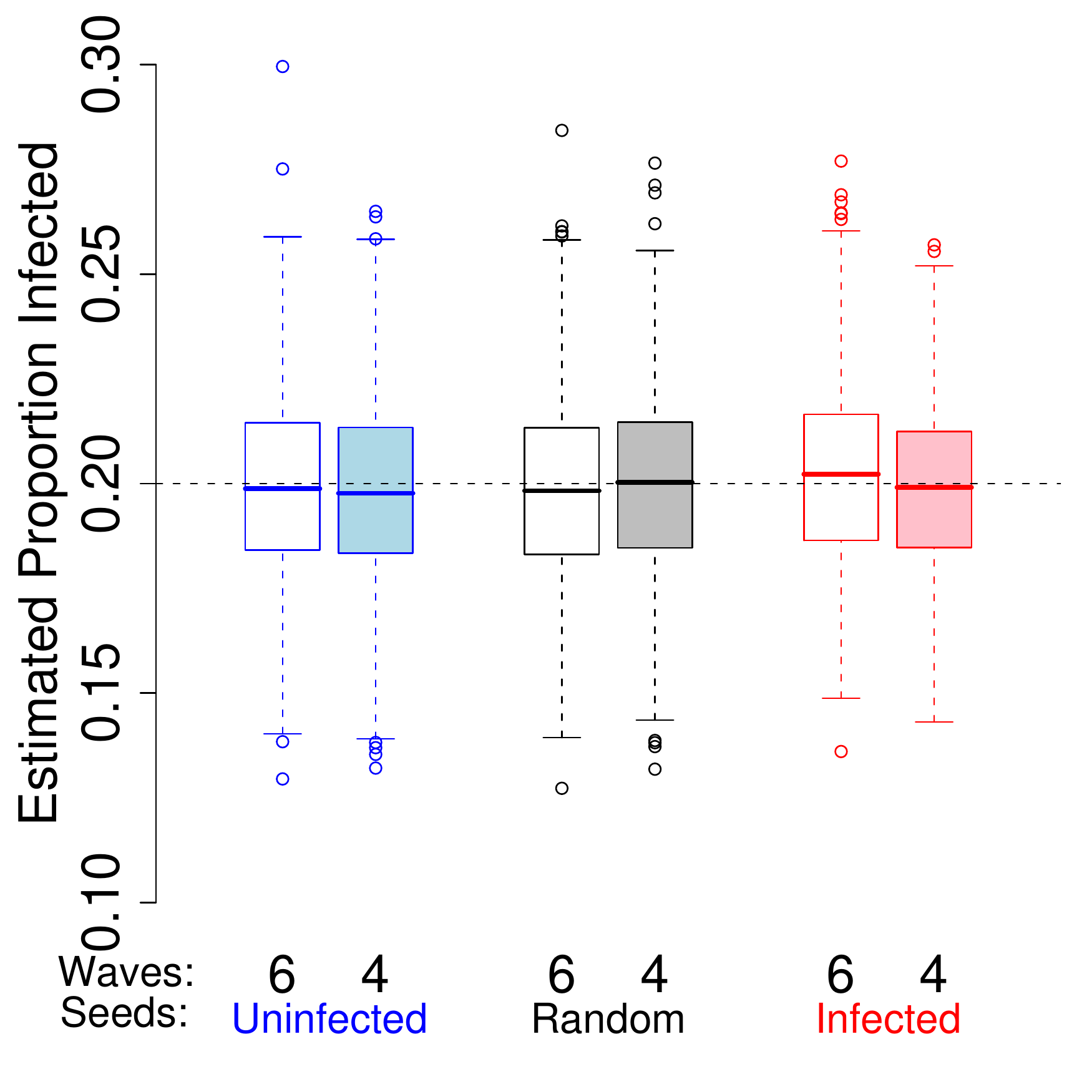}
}\vspace{.5cm}
\subfigure[Discard Seeds, Wave 1, 2]
{
    \label{withouts12}
    \includegraphics[width=\sss,height=\ttt]{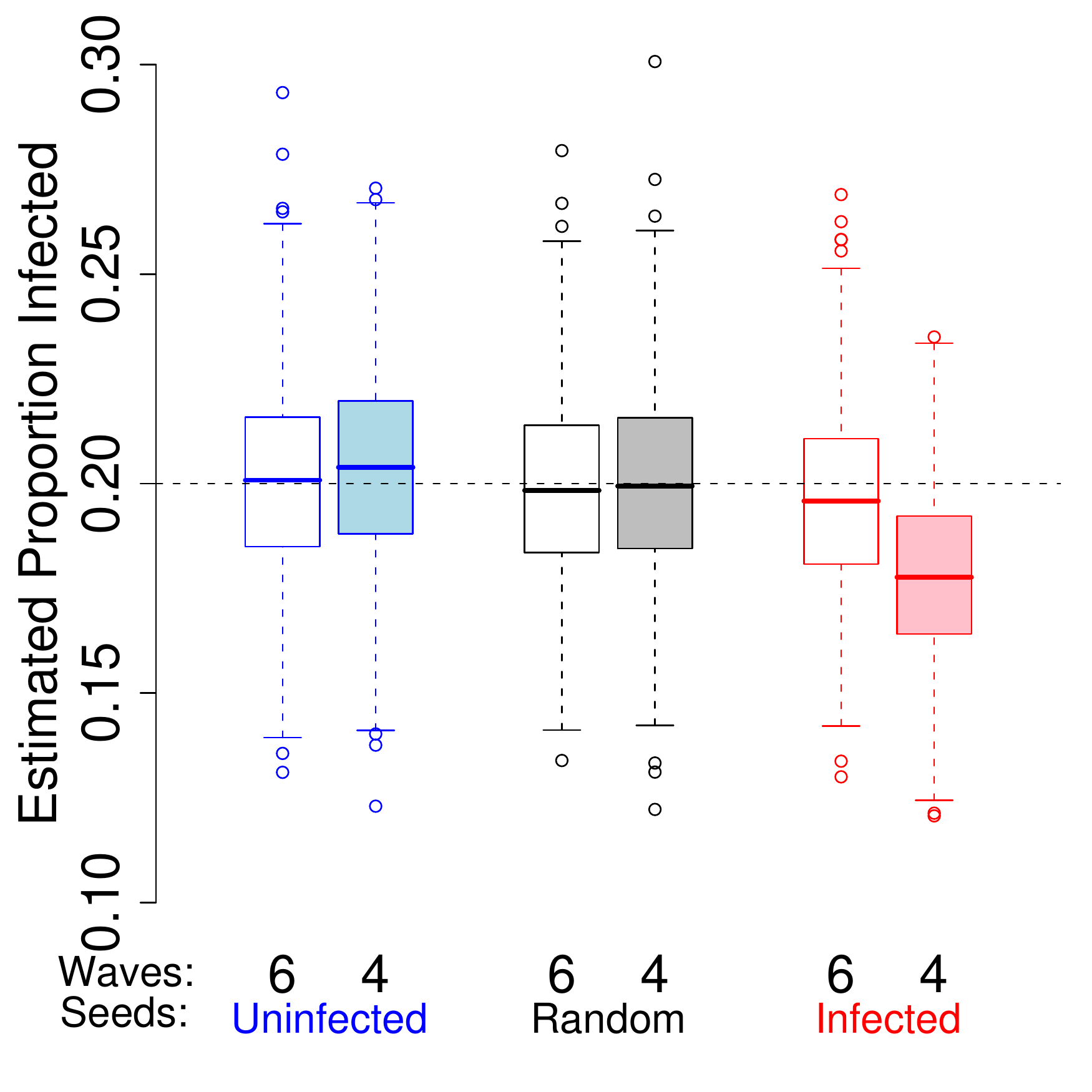}
}\hspace{.5cm}
\subfigure[Discard Seeds, Wave 1,2,3]
{
    \label{withouts123}
    \includegraphics[width=\sss,height=\ttt]{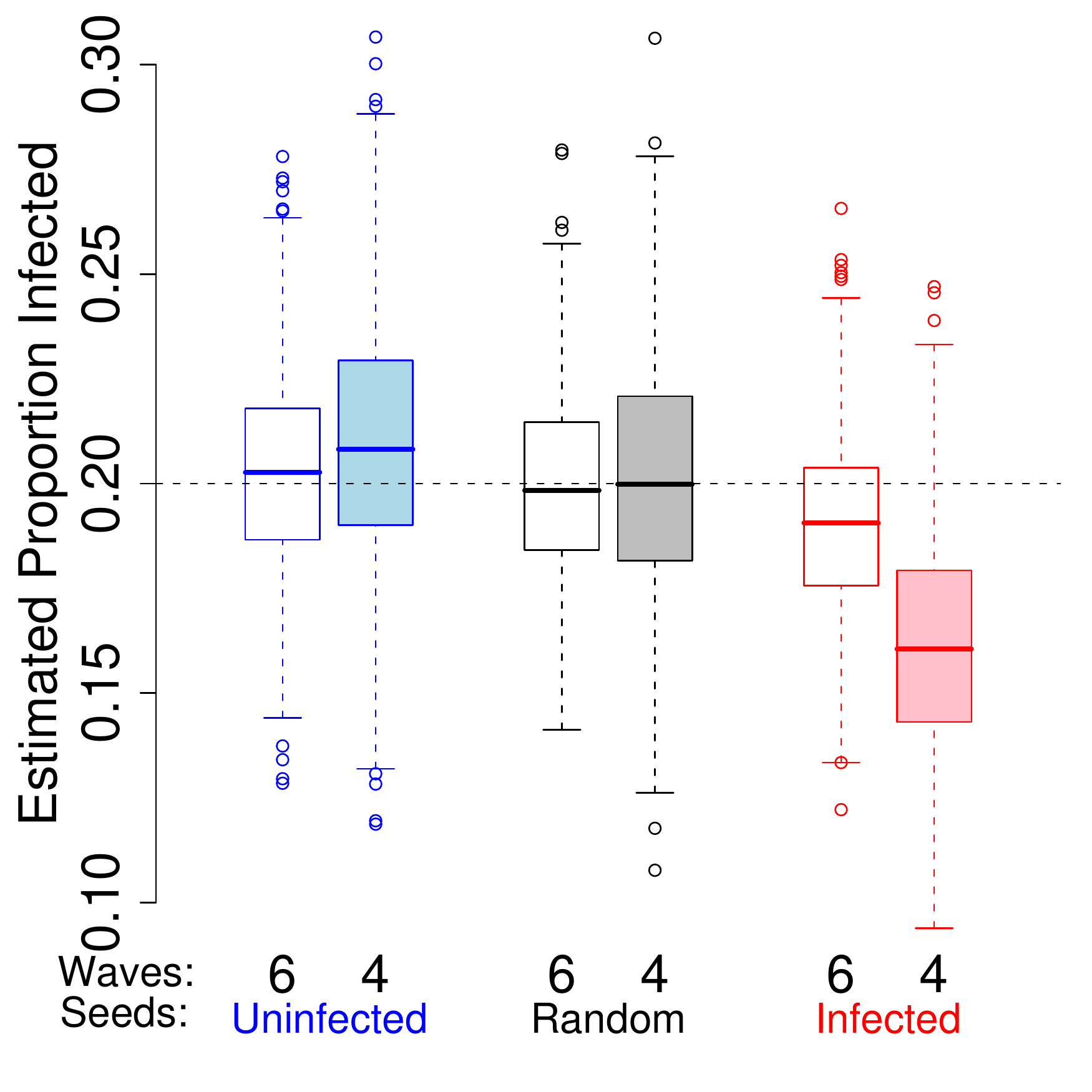}
}
\end{center}
\caption[Comparison of V-H Estimators with Early Waves Discarded]{V-H estimators from samples of 6 seeds, 6 waves (first, third, and fifth boxes), and 20 seeds, 4 waves, from seeds selected from all uninfected nodes (first two boxes), random nodes with respect to infection (second two boxes), and all infected nodes (last two boxes).  Seeds are discarded from all estimators.  Additional early waves discarded from second, third, and forth plots.}  \label{withoutlots}
\end{figure}

As in Figure \ref{cdcwave}, Figure \ref{withouts}, discarding only the seeds, illustrates bias in the conditions with biased seeds, with more substantial bias in the simulations with fewer waves (the second and sixth boxes).  Figure \ref{withouts1} depicts estimators based on the same samples, this time discarding the first wave of samples along with the seeds.  This plot illustrates the motivation for burn-in in MCMC processes, and provides support for the notion of discarding earlier waves to reduce the bias induced by seeds in RDS analysis.  Comparing Figure \ref{withouts} and Figure \ref{withouts1}, the bias in the estimators virtually disappears, with little impact on the variance.

Figure \ref{withouts12}, however, demonstrates a potentially problematic feature of this strategy.  When two waves are discarded, bias returns to the estimator, but in the other direction.  The samples beginning with all infected nodes now exhibit negative bias, and to a greater degree in the condition with fewer waves. This effect is exacerbated when three waves, in addition to the seeds, are discarded, as in Figure \ref{withouts123}.  Here, the bias in the infection-seeded samples is substantially larger, and positive bias is also apparent in the uninfected-seeded samples with fewer waves.

This bias-inducing impact of discarding early samples is due to the without-replacement nature of the sampling process.  In a true with-replacement random walk process,  discarding early waves would continue to decrease seed bias.  In the RDS sampling process, however, nodes that are sampled once in discarded waves cannot be re-sampled.  Consider the case of infected seeds and {\it few waves} of sampling.  There are 20 seeds, all infected.  Consider the extreme case of perfect homophily.  Then the first wave would consist of 40 additional infected nodes.  The second wave would consist of an additional 80 infected nodes.  Discarding the seeds and first two waves would remove 130 of the population total 200 infected nodes from the possibility of inclusion in the estimator.  For this reason, although early waves are more dependent on seeds than later waves, it is unclear under which circumstances discarding early waves will improve performance of the estimator, and under which conditions it will decrease performance.  An alternative approach would be to estimate the relative inclusion probabilities of all sampled nodes conditional on the composition of the seeds, and compute estimators based on those probabilities.  An approach of this sort is presented in \cite{gile08} and \cite{gilehan09rds}.

\subsection{Respondent Behavior:  Random Referral} \label{section:bias}

We consider the effects of respondent behavior in selecting alters for referral on the V-H estimator.  For this comparison, we compare the standard condition of coupon distribution completely at random among eligible alters to the condition where an infected alter is 20\% more likely to be sampled than an uninfected.

\begin{figure}
\begin{center}
    \includegraphics[width=4.5in,height=4.5in]{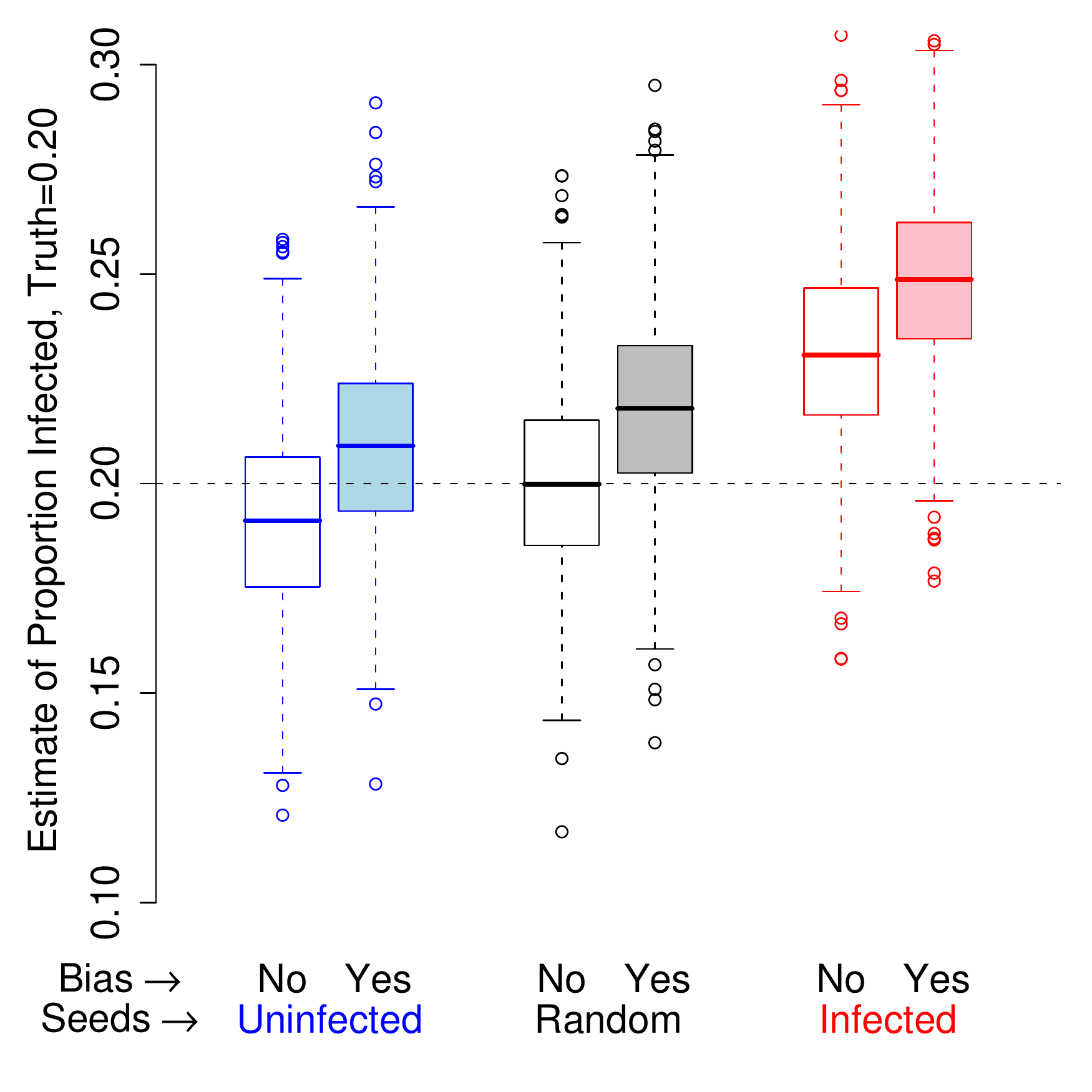}
\end{center}
\caption[Comparison of V-H Estimators with Biased and Unbiased Referral]{V-H estimators from samples with unbiased referral (first, third, and fifth boxes), and referral 20\% more likely for infected partners, from seeds selected from all uninfected nodes (first two boxes), random nodes with respect to infection (second two boxes), and all infected nodes (last two boxes).}  \label{cdcbias}
\end{figure}

The results of this study are depicted in Figure \ref{cdcbias}.  We see that under all three sets of seed-selection conditions, the simulations with biased referral result in increased positive bias in the estimator.  This is not surprising, as in this case the infected nodes are more likely to be included in the sample, while this bias is not accounted for in parameter estimation.  The magnitude of the increased bias is about the same across the three seed-selection regimes.

\subsection{Random Walk Model}
In this section, we focus on the with-replacement assumption of the random walk model.  \cite{volzheck08} argue that this assumption is close to accurate for small sample fractions (or populations much larger than the sample size).  For this reason, in the first sub-study, we compare the performance of the estimator with varying sample fractions.  In the second sub-study, we compare the performance of the estimator under with and without replacement sampling.

\subsubsection{Network Size Large (compared to sample, $N >> n$)}\label{section:simsmall}

\begin{figure}
\begin{center}
    \includegraphics[width=2in]{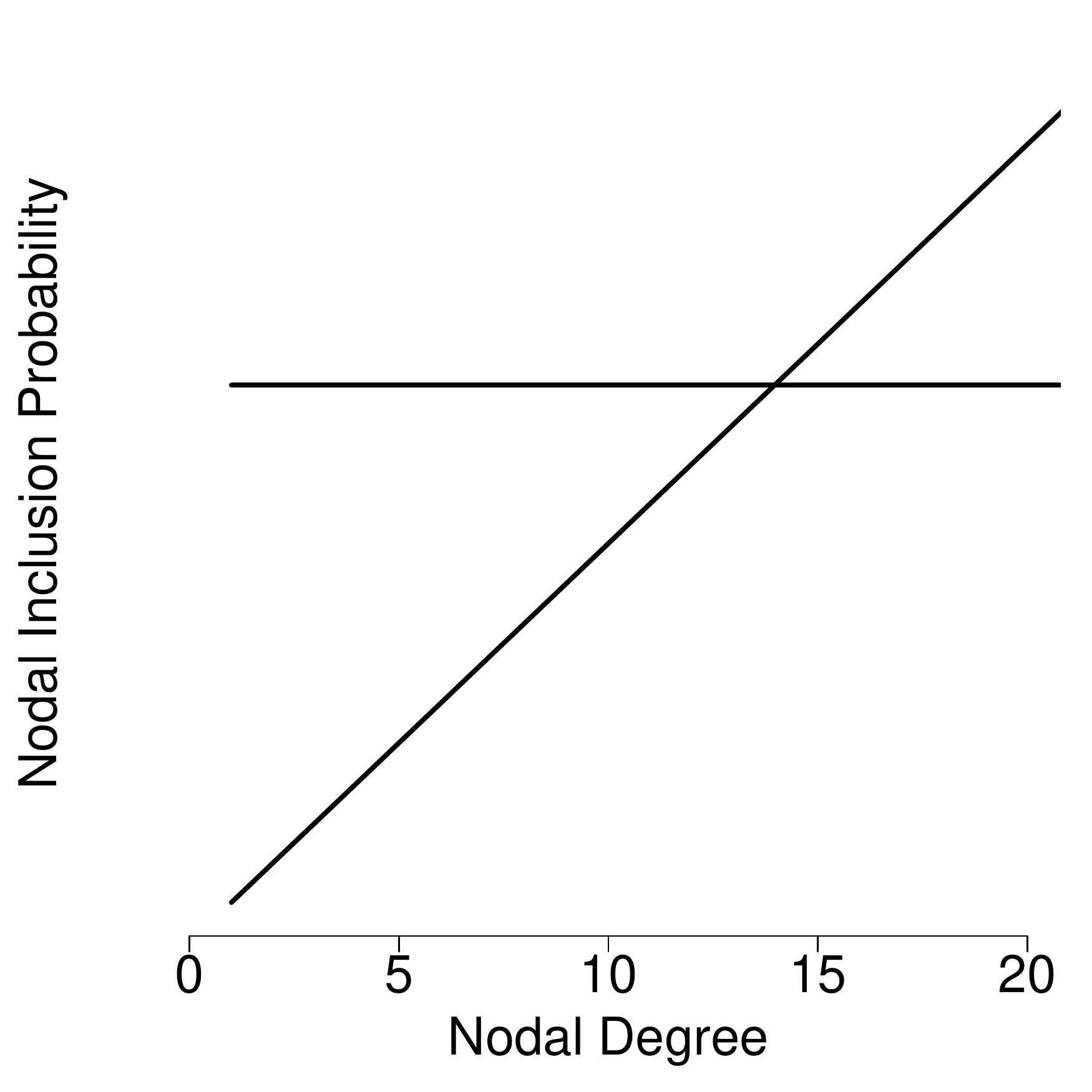}
\end{center}
\caption{Heuristic depiction of the mapping from nodal degree to sampling probability for full population sample (horizontal line) and random walk model (diagonal line).}  \label{twolines}

\end{figure}

The use of sampling probabilities from a with-replacement process in a without-replacement context is often justified by a small sample proportion.  If the sample does not substantially deplete the population, then the sampling probabilities of the remaining units are nearly unchanged by the removal of the sample.  In many RDS studies, however, the sample is known to include a large fraction of the target population \citep{Malekinejad:2008ty}. \knote{check this cite}
In the extremal case of sampling the full population, far from being proportional to degree, the inclusion probabilities are all equal to 1.  Figure \ref{twolines} illustrates the relationship between nodal degree and sampling probability for the random walk model (diagonal line, representing probability proportional to degree) and the case of a full population sample, in which all sampling probabilities are equal (horizontal line).  The purpose of this section is to investigate the impact of large sample fractions on the performance of the V-H estimator.

In the case of RDS, the net advantage of a small sample fraction is more ambiguous than in standard survey sampling due to the competing interest in reaching the equilibrium distribution of the random walk.  Although this is unreasonable to expect on the nodal level,
it may be more reasonable on the level of mixing of nodal classes.    Furthermore, since interest is focused on estimating population proportions, the relative sampling weights of individuals may be quite inaccurate, but not bias the estimators of interest.

The greater the sample proportion, the greater the deviation from the probability proportional to degree sampling assumption.  This type of inaccuracy does not induce bias in estimators unless it differentially effects infected and uninfected nodes.  For communicable diseases, it is typical to find different patterns of social relations between infected and uninfected persons.  \cite{frost06}, for example, find that male injection drug users in Tijuana testing negative for Syphilis have about 1.7 times as many contacts as their positive-testing counterparts.
For this reason, in this series of simulations, we evaluate the relationship between sample fraction and relative activity level of infected and uninfected nodes.  In particular, we control the ratio
\[
w = \frac{{\bar d}_I}{{\bar d}_U}
\]
where ${\bar d}_I$ is the mean degree of the infected nodes and ${\bar d}_U$ is the mean degree of the uninfected nodes.  We consider cases where $w$ takes the values 1, 1.1, 1.4, 1.8, and 3.  In each case, the mean degree ${\bar d}$ of the network remains fixed at 7.

To avoid confounding the effects of sample size with the effects of sample proportion, we fixed the sample size at 500 nodes, while treating populations of sizes 1000, 835, 715, 625, 555, and 525 resulting in sample proportions of about 50\%, 60\%, 70\%, 80\%, 90\%, and 95\%.  For this part of the study, all seeds are chosen at random with respect to infection status.

\begin{figure}
\begin{center}
    \includegraphics[width=4in,height=4in]{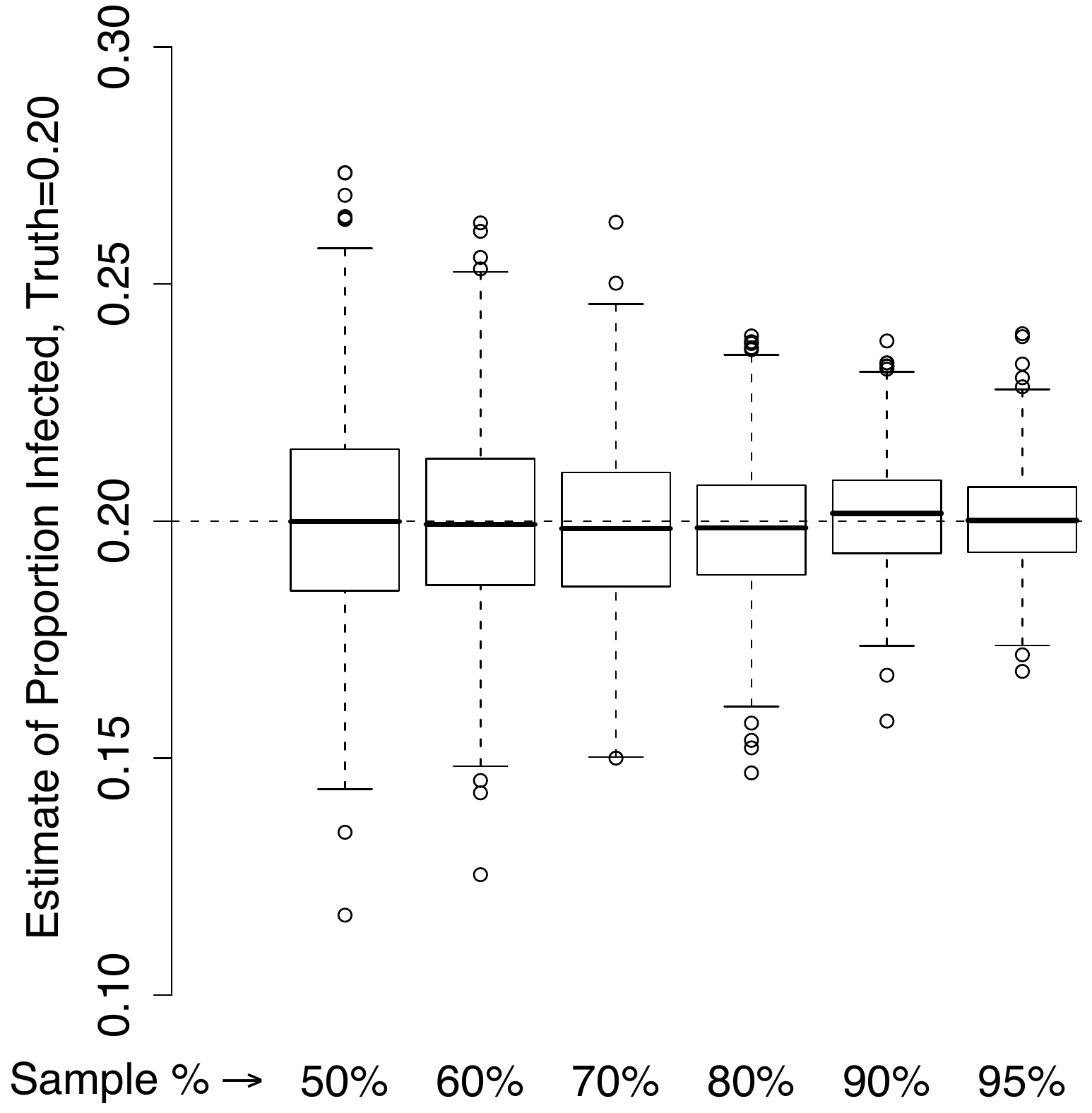}
\end{center}
\caption[Comparison of V-H Estimators with Varying Sample Fraction]{Current RDS estimators from samples of size 500 constituting about  50\%, 60\%, 70\%, 80\%, 90\%, and 95\% of the population.  All seeds selected with probability proportional to degree independent of infection status.  Infected and uninfected nodes equally active ($w=1$). }  \label{small1}
\end{figure}

The first plot in this series, Figure \ref{small1}, depicts the standard case in which the two groups are equally active ($w=1$).  In this case, there is negligible bias
in the estimators for any of the population sizes.
Also, not surprisingly, the variance of the estimators is reduced as the sample fraction increases.  If 95\% of the population is sampled, there is little room for variance in the estimator.

\newcommand{\corn}{2in}
\newcommand{\dog}{2in}

\begin{figure}[h]
\begin{center}
\subfigure[$w=1.1$]
{
    \label{small1.1}
    \includegraphics[width=\corn,height=\dog]{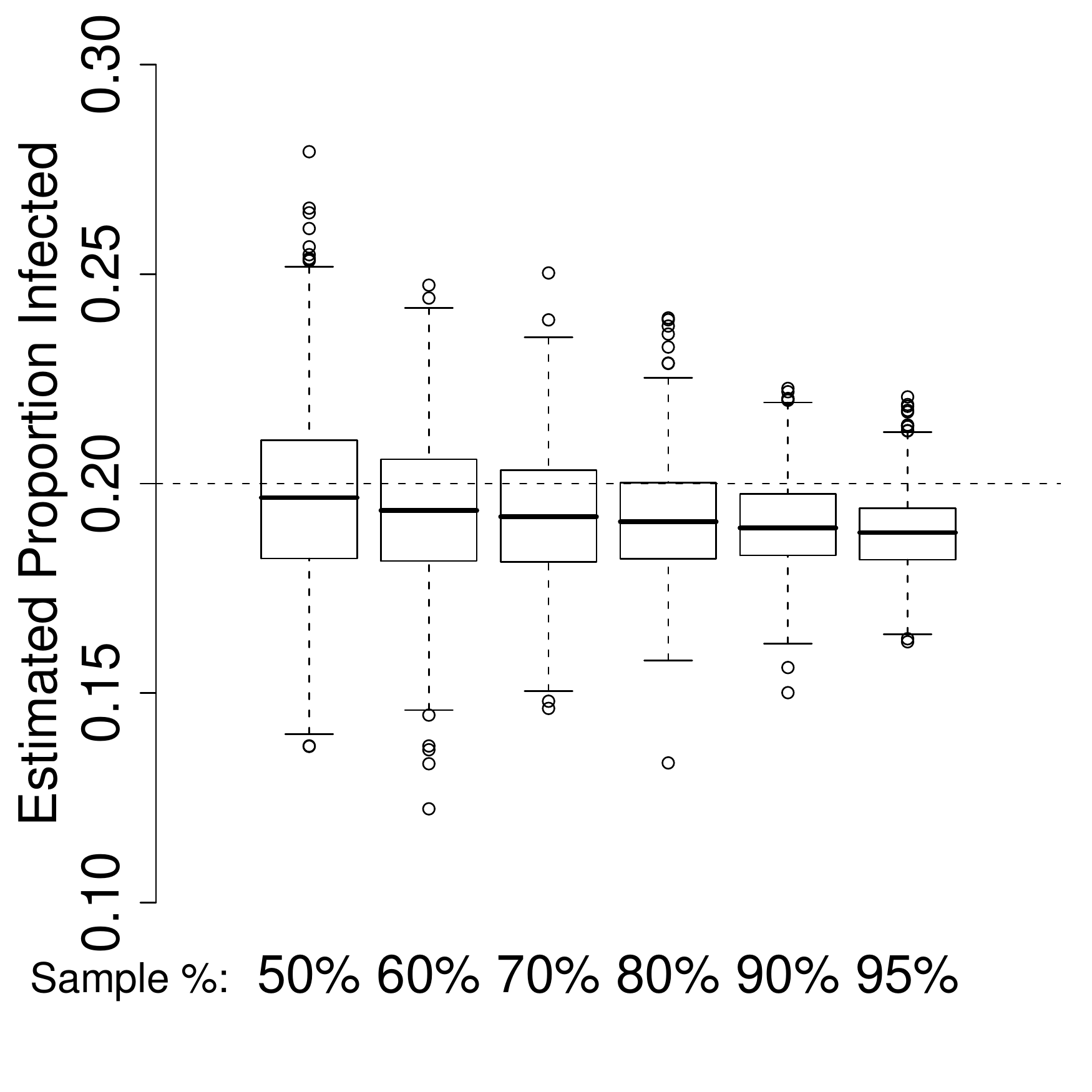}
} \hspace{.5cm}
\subfigure[$w=1.4$]
{
    \label{small1.5}
    \includegraphics[width=\corn,height=\dog]{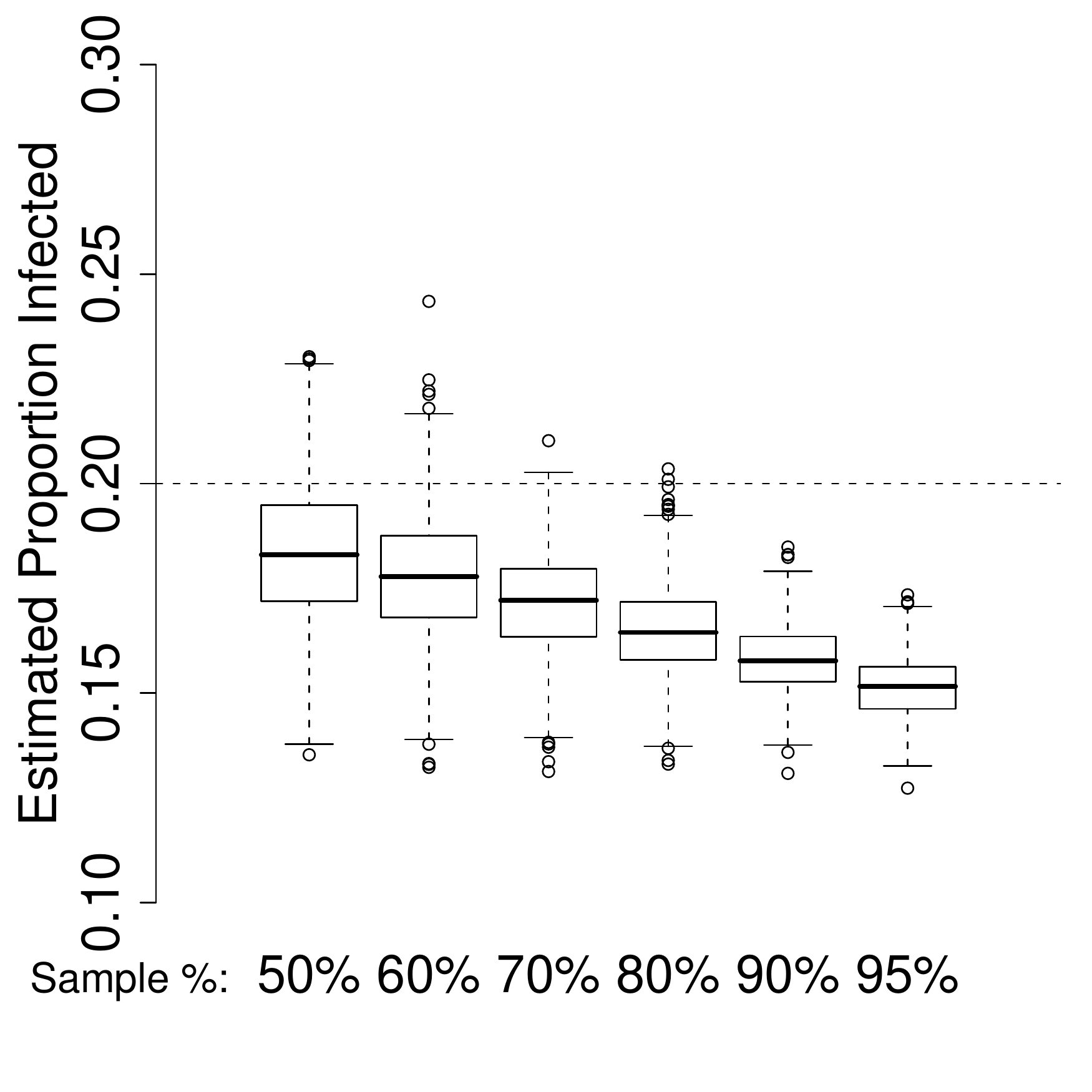}
}\vspace{.5cm}
\subfigure[$w=1.8$]
{
    \label{small2}
    \includegraphics[width=\corn,height=\dog]{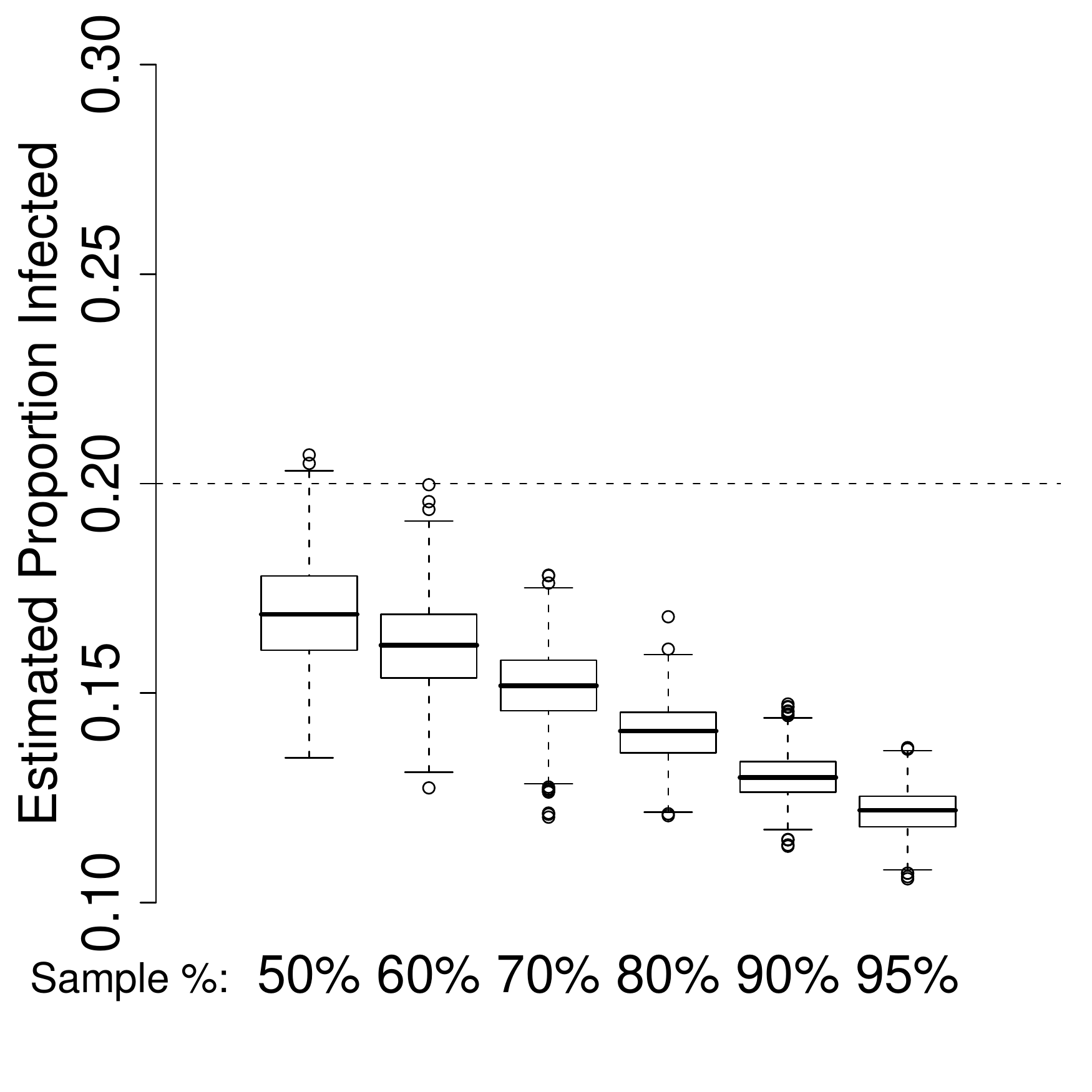}
}\hspace{.5cm}
\subfigure[$w=3$]
{
    \label{small4}
    \includegraphics[width=\corn,height=\dog]{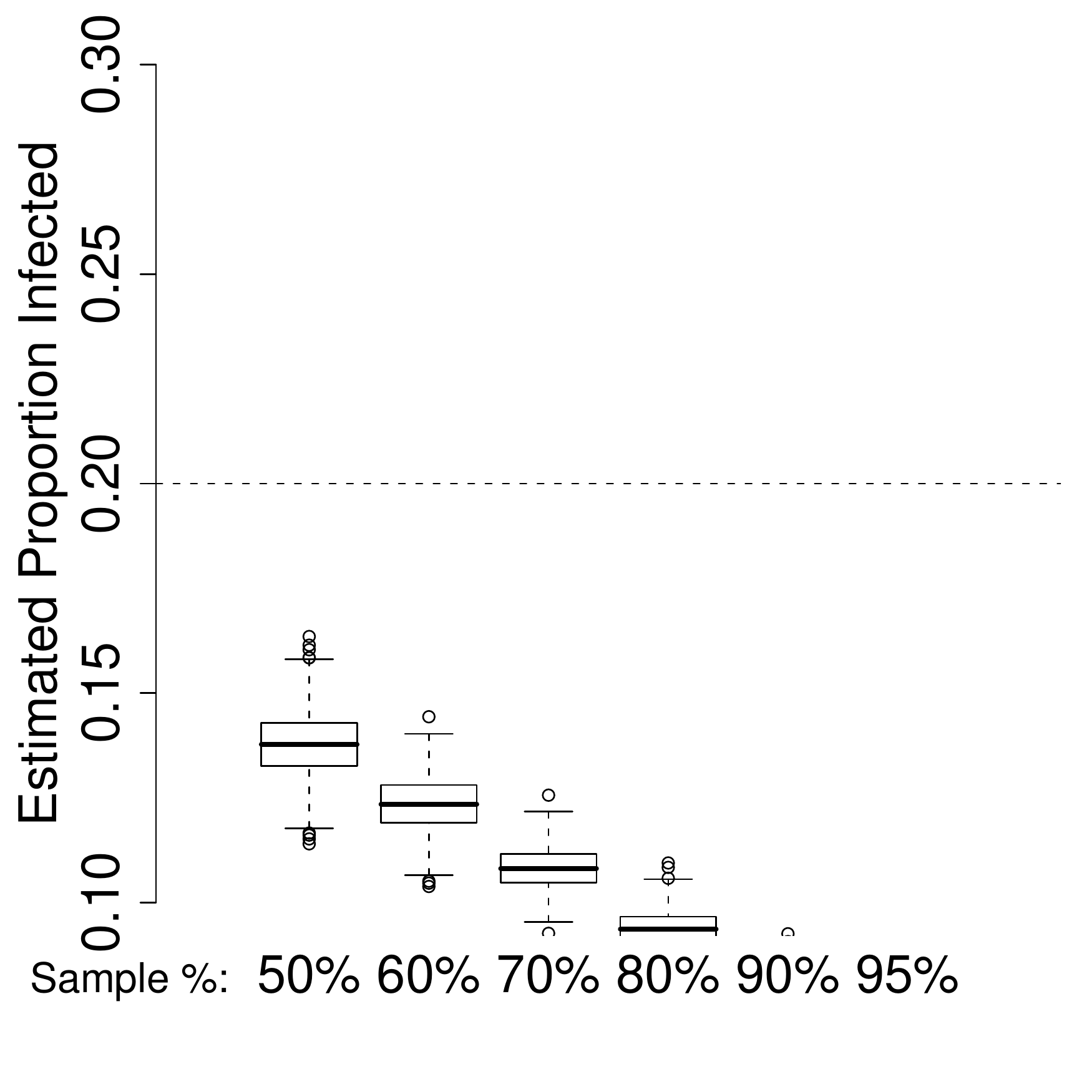}
}
\end{center}
\caption[Comparison of V-H Estimators with Varying Sample Fraction and Varying Relative Activity Level]{V-H estimators from samples of size 500 constituting about  50\%, 60\%, 70\%, 80\%, 90\%, and 95\% of the population.  All seeds selected with probability proportional to degree independent of infection status.  Subfigures with varying degrees of elevated activity of infected nodes ($w$).} \label{smalllots}
\end{figure}

When the relative activity level $w$ increases only a little, to 1.1, a substantial bias is evident in Figure \ref{small1.1}, which persists even as the sample proportion increases to 95\%.  This is because as the sample fraction increases, the true inclusion probabilities become closer to uniform (horizontal line in Figure \ref{twolines}).  In the estimator, however, the higher-degree nodes are down weighted proportional to their degree (diagonal line Figure \ref{twolines}). When true inclusion probabilities are closer to uniform, this constitutes over-weighting low-degree nodes and under-weighting high-degree nodes.  When the infected nodes are disproportionately of high degree, this constitutes under-representation of that group, and induces the observed negative bias.

\begin{figure}[h]
\begin{center}
    \includegraphics[width=5in]{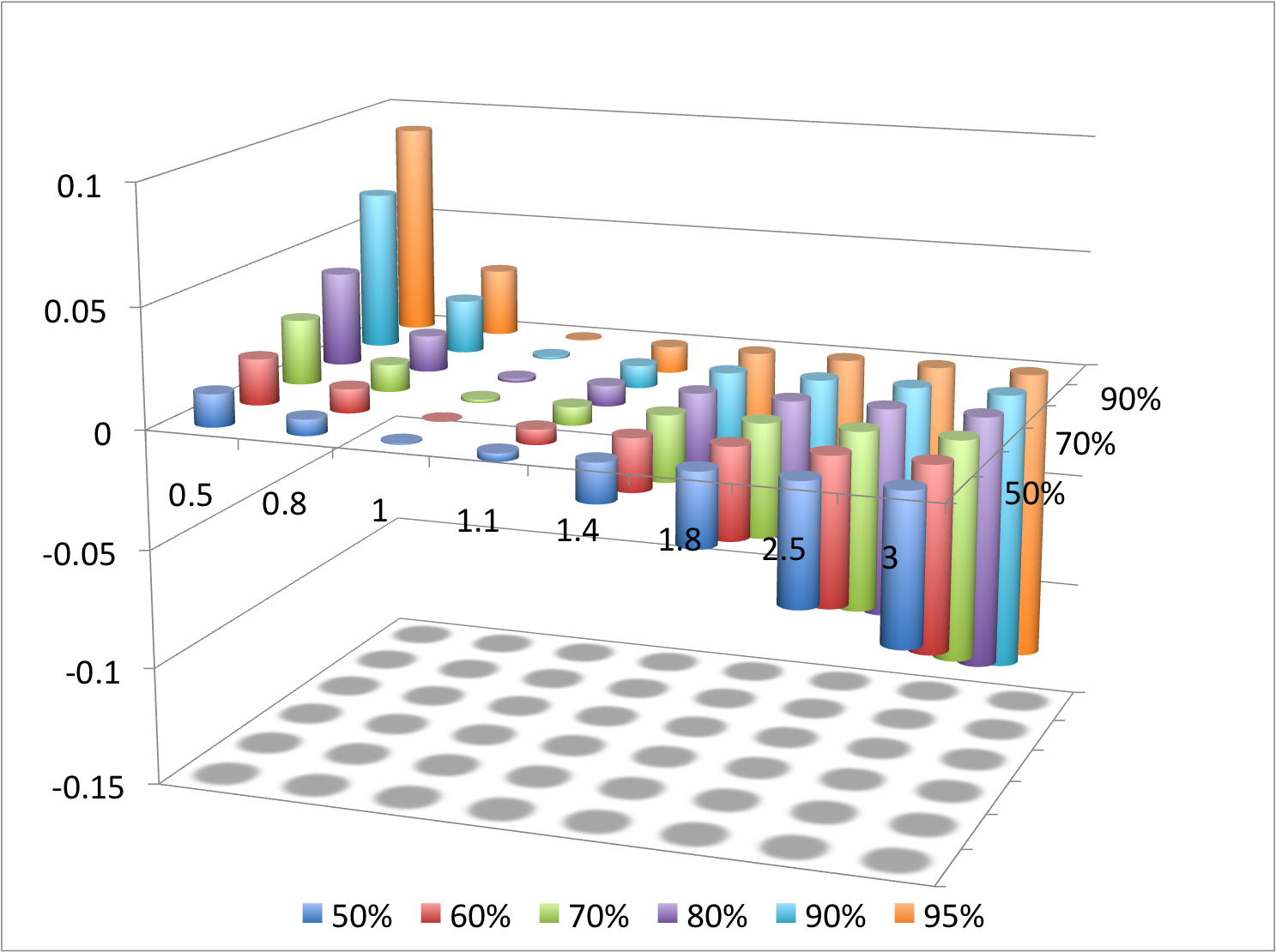}
\end{center}
\caption[Bias of the V-H estimator Varying Sample Fraction and Varying Relative Activity Level]{Bias of the Current RDS estimator from samples of size 500 constituting about  50\%, 60\%, 70\%, 80\%, 90\%, and 95\% of the population, for varying degrees of elevated activity of infected nodes ($w$).} \label{fig:biasbars}
\end{figure}

Figures \ref{small1.5}, \ref{small2}, and \ref{small4} illustrate the increase in this bias as the relative activity level, $w$, of the infected nodes increases to 1.5, 1.8, and 3, respectively.  As the relative activity level of the infected nodes increases, the negative bias increases as well.  The absolute bias is also larger for larger sampling fractions.  When infected nodes are eighty percent more active than uninfected nodes ($w=1.8$), the negative bias is so strong that only a few of the estimators are above the true value, and all of these in the smallest sample fraction condition.  When $w=3$, even the largest estimators are more than .03
below the true value for the 50\% sample condition, a distance of more than 4.5 standard deviations.  In the 95\% sample condition, even the most extreme simulated samples estimate the population proportion at less than half its true value.  Figure \ref{fig:biasbars} allows for the side-by-side comparison of the biases induced in the V-H estimator for sample percentages 50\% through 95\% and values  of $w$ between 1 and 3, as well as below 1 (i.e. infected nodes less active than uninfected).  This figure illustrates that the V-H estimator is biased for relative activity level $w=1$, then has increasing negative bias for larger values of $w$ and increasing positive bias for smaller fractional values of $w$, with these effects exacerbated for larger sample fractions.

 It is also interesting to note that in all these simulations, the variance of the estimators decreases with sample fraction.  This is to be expected, as the number of possible nodal samples drops
from $1000 \choose{500}$ for the 50\% samples to $525 \choose{500}$ for the 95\% samples.  Note that because of the greater variance in degree in the simulations with higher $w$, the sample composition is even less variable, leading to lower variance.

Because the bias in these samples is due to a known inaccuracy of sampling weights, it would seem to be a correctable error.  One such correction is presented in \cite{gile08} and \cite{gilehan09rds}.

We also note that the simulations in Figures \ref{small1}, \ref{smalllots}, and \ref{fig:biasbars} were repeated with varying degrees of homophily by infection status, as well as with clustering on an orthogonal nodal variable, with no substantial differences from the results in Figures \ref{small1} and \ref{smalllots}.

\subsubsection{With-Replacement Sampling} \label{asssecwith}

In this section we explore the performance of current RDS estimation if sampling were conducted with replacement.  Practically, this might constitute allowing persons to participate in the sample more than once.  Recall that the current RDS estimator is theoretically justified by a with-replacement approximation.  For this reason, we might expect the estimator to demonstrate superior performance when sampling is with replacement.  All previous studies of the performance of the RDS estimators have accepted the with replacement approximation.

\begin{figure}[h]
\begin{center}
    \includegraphics[width=4.5in,height=4.5in]{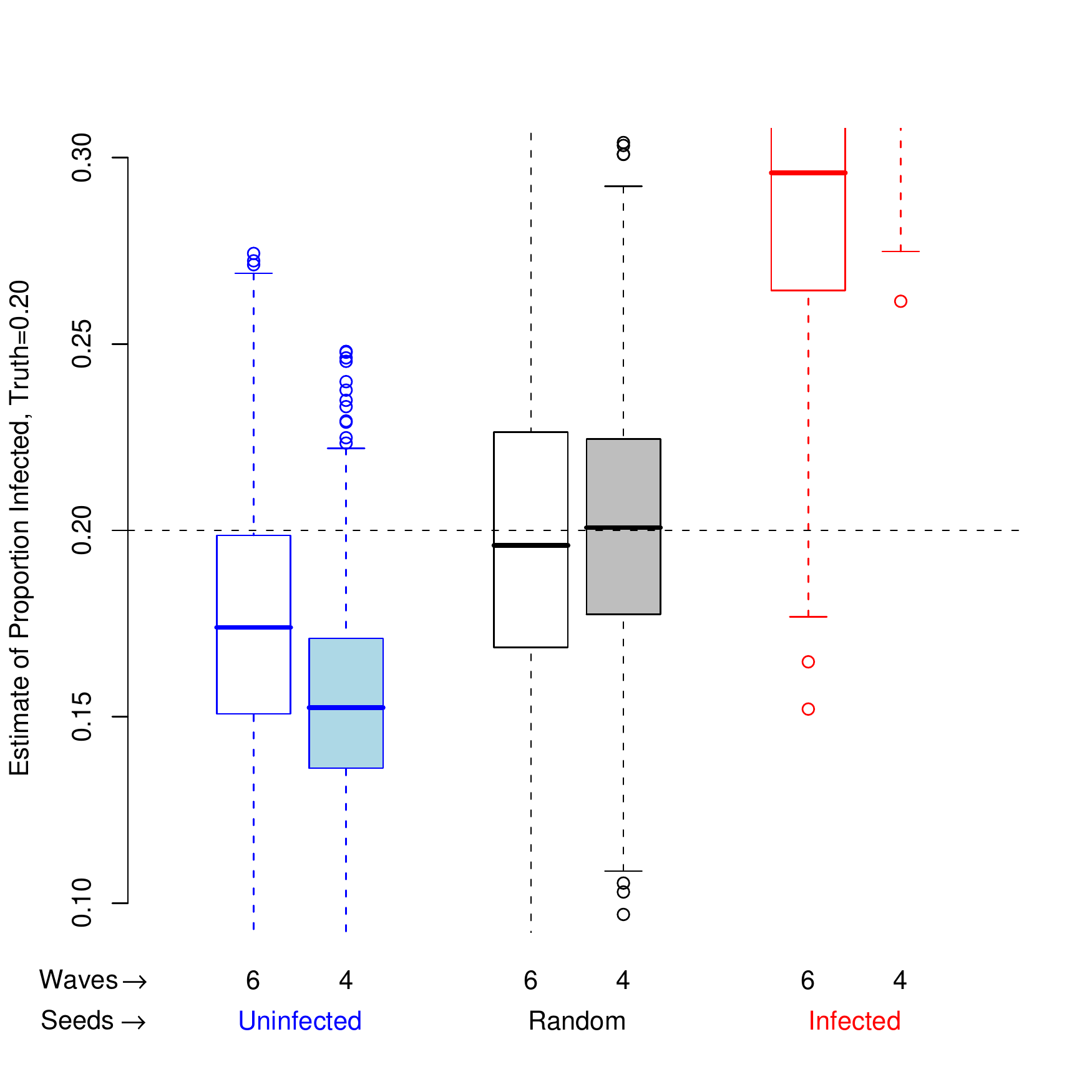}
\end{center}
\caption[Comparison of V-H Estimators with Many and Fewer Sampling Waves, Sampling With Replacement]{V-H estimators from samples with replacement consisting of 6 seeds, 6 waves (first, third, and fifth boxes), and 20 seeds, 4 waves, from seeds selected from all uninfected nodes (first two boxes), random nodes with respect to infection (second two boxes), and all infected nodes (last two boxes).} \label{cdcwavewith}
\end{figure}

In this sub-study, we replicate the sub-study in Section \ref{section:waves}, only simulating coupons passed to alters completely at random, regardless of whether the alter was already in the sample.  In this case, some nodes are included in the estimator multiple times.  As in Section \ref{section:waves},  samples are taken with more and fewer waves (6 seeds, 6 waves, and 20 seeds, 4 waves, respectively), and with each of the three types of seed selection.  The distributions of the resulting estimators are summarized in Figure \ref{cdcwavewith}.

The bias and variance of the resulting estimators are both far greater than the corresponding without-replacement results (Figure \ref{cdcwave}).
The increased variance can be explained by the larger space of possible samples, with $1000^{500}$ possible samples in the with-replacement case, as compared to only $1000 \choose{500}$ possible samples in the without replacement case.

Figure \ref{cdcwavewith} also illustrates far greater bias in the direction of the seed bias, for all the cases with biased seeds.  The critical point is that without-replacement sampling encourages the faster mixing of the sampling process.  In the with-replacement case, however, a sample seeded by infected nodes can continue to sample infected nodes indefinitely.  Remarkably, this suggests that violation of the assumption of with-replacement sampling results in improved performance of the estimator.

\section{Comparison of the Classic and Current RDS Estimators}\label{sec:sh}

Sections \ref{sec:rdsest} and \ref{sec:sh} focus on the
current RDS estimator, proposed by \cite{volzheck08} (V-H).
This estimator is also more closely tied to previous work in survey sampling than the
previous estimators. As noted in Section \ref{sec:rdsest},
most RDS analysis has been conducted using the classic estimator
proposed by \cite{salgheck04} (S-H), and
implemented in the standard RDS software, RDSAT \citep{rdsat}.

In this section we compare the
properties of these two estimators.  We first introduce the
classic estimator, then compare the performance of the two
estimators in the simulation studies presented in the preceding
section.  We conclude that the V-H estimator consistently
out-performs the S-H estimator.

We consider the classic RDS estimator \citep{salgheck04} in the case of estimating the population proportion of group $A$ in a population partitioned into groups $A$ and $B$.  This estimator is based on equating the estimated number of relations from $A$ to $B$ to the estimated number of relations from $B$ to $A$.  This is done by first estimating the mean degree of each group ($d_A$ and $d_B$, respectively), then by estimating the proportion of relations of group $A$ which are relations to group $B$, and visa versa.  So if $N_A$ is the population size of group $A$, and $N_B$ the population size of group $B$, $N_A+N_B=N$, then let $t_{AB}$ represent the total number of relations between groups $A$ and $B$. Then the proportion of each group's relations that are shared with the other group are defined by:
\begin{align}
c_{AB}&= \frac{t_{AB}}{N_A{\cdot}d_A} \\
c_{BA}&= \frac{t_{AB}}{N_B{\cdot}d_B}.
\end{align}
Algebraic manipulation results in the form of the classic RDS estimator:
\bea
\mu_A = \frac{N_A}{N}= \frac{d_B \cdot c_{BA}}{d_A \cdot c_{AB} + d_B \cdot c_{BA}}.
\label{shform}
\eea
\cite{salgheck04} then estimate the unknown mean degrees $d_A$ and $d_B$ using generalized Horvitz-Thompson estimators assuming nodal sampling probabilities proportional to degrees.  Note that they motivate this estimator based on the random walk model which is central to the current RDS estimator.  Therefore, they rely on the set of assumptions in Table \ref{tab:assred}.
 They then estimate the unknown proportions of cross-group relations as the observed proportions of cross-group referrals.  For this estimate, they rely on the fact that in a converged random walk, edges are sampled with equal probability.  It is unclear how biases resulting from this assumption compare to those induced by the nodal sampling assumptions of the V-H estimator.  \cite{salgheck04} substitute both sets of estimates into (\ref{shform}) to form their estimator ${\hat\mu}_A$.

Thus, \cite{salgheck04} also rely on a large set of assumptions, some of which are known to be approximations.  In addition, the general form of their estimator involves substituting estimates into equations in a somewhat {\it ad hoc} manner.  For this reason, and because we expect the more principled current estimator to supplant the classic estimator, we do not address the properties of the classic estimator at length.  Instead, we present a brief comparison of the classic and current estimators for the simulation studies considered in this paper.  Tables \ref{shvhfirst} and \ref{shvhsmall} present the primary results of this comparison.  Each entry in these tables represents the relative efficiency of the classic estimator as compared to the current estimator, computed as the ratio of mean squared errors:
\bea
\textrm{Relative efficiency} = \frac{MSE_{\kern1pt\textrm{V-H}}}{MSE_{\kern1pt\textrm{S-H}}}.
\eea
Nearly all of these relative efficiencies are below 1, indicating the superior performance of the V-H estimator.  In nearly all of these cases, the V-H estimator also has both lower bias and lower variance than the S-H estimator.  It is also interesting to note that the bias induced by seed composition in the S-H estimator is in the opposite direction to that of the V-H estimator.  Consider the case of all infected seeds, in which infected nodes are systematically over-sampled.  Due to the finite population, fewer infected nodes are available for recruitment.  Because the S-H estimator relies on the proportion of within-group and cross-group referrals to estimate the proportion of cross-group relations, this results in a systematic under-estimation of the size of the infected sub-population.  In fact, in the empirical comparison of these two estimators provided by \cite{volzheck08}, the race and gender estimates provided by these two estimators are on opposite sides of the sample mean, suggesting that the seed selection bias on these two variables may not have been removed.

A notable exception to the superior performance of the V-H estimator is provided by the case of biased referral (infected referred more often) and all infected seeds, the last entry in Table \ref{shvhfirst}.  The relative efficiency here is $2.72$.  The apparently superior performance of the S-H estimator in this case is because the negative bias induced by the seed selection is partly corrected by the bias induced by the referral pattern.

\begin{table}[h]\caption{Relative efficiency of the S-H to the V-H estimator, for simulation in Sections \ref{section:waves}, \ref{section:homoph}, and \ref{section:bias}}
\begin{center}
\begin{tabular}{l||cc|cc|cc}
Seeds & \multicolumn{2}{c}{\blue Uninfected} & \multicolumn{2}{c}{Random} & \multicolumn{2}{c}{\red Infected}\\
\hline
\hline
Waves & 6 & 4 & 6 & 4 & 6 & 4 \\
\hline
(Section \ref{section:waves})& 0.94 & 0.84 & 0.78 & 0.73 & 0.88 & 0.82 \\\\
\hline
\hline
Homophily & Low & High & Low & High & Low & High \\
\hline
(Section \ref{section:homoph})& 0.97 & 0.83 & 0.72 & 0.56 & 0.95 & 1.23 \\\\
\hline
\hline
Referral Bias & No & Yes & No & Yes & No & Yes \\
\hline
(Section \ref{section:bias})& 0.97 & 1.07 & 0.72 & 0.42 & 0.95 & 2.72 \\
\end{tabular} \label{shvhfirst}
\end{center}
\end{table}

\begin{table}[h]\caption{Relative efficiency of S-H to the V-H estimator, for simulation in Section \ref{section:simsmall}}
\begin{center}
\begin{tabular}{c||cccccc}
$w$ & 50\% & 60\% & 70\% & 80\% & 90\% & 95\% \\
\hline
\hline
1 & 0.97 & 0.77 & 0.54 & 0.42 & 0.24 & 0.20 \\
  1.1 & 0.83 & 0.78 & 0.61 & 0.40 & 0.35 & 0.30 \\
  1.4 & 0.60 & 0.53 & 0.48 & 0.47 & 0.50 & 0.52 \\
  1.8 & 0.54 & 0.54 & 0.56 & 0.56 & 0.59 & 0.62 \\
  3 & 0.70 & 0.72 & 0.75 & 0.77 & 0.79 & 0.80 \\
\end{tabular} \label{shvhsmall}
\end{center}
\end{table}

In summary, we find that the Volz-Heckathorn estimator out-performs the Salganik-Heckathorn estimator in almost all circumstances.  In addition the Volz-Heckathorn estimator is also easier to compute, and applies directly to continuous as well as categorical variables.

\knote{throughout:  do I want to talk about three sources of bias, or about three types of sensitivity or three types of assumptions, or something else?}
\mnote{I would vote not to do any more in this version. It reads well to me. We will get another shot in the revision (with luck) if you do not like this discussion.}

\section{Discussion and Recommendations} \label{sec:ch5discuss}
RDS methodology addresses an under-served need by allowing for something like probability sampling in the context of hidden populations.  However the performance of current estimators is
sensitive to deviations from ideal sampling conditions in three areas:  convenience sampling of seeds, respondent behavior, and deviations from the random walk model.
We structure our discussion of current RDS methodology around these three areas.

\subsection{Bias Induced by Seed Selection}
The main advantage of RDS methodology over time-location sampling or non-probability methods for sampling hidden populations is that the long sampling chains reduce, or ideally eliminate, the biases induced by the convenience sampling of seeds.  If seeds were selected completely at random, or with some other know probability distribution such as with probability proportional to degree, then RDS would not be necessary; a sufficiently sized sample of seeds would allow for estimation of any population proportion of interest.  Previous research has relied on an ambiguous notion of {\ql}asymptotic unbiasedness{\qr} to support the use of the V-H and S-H estimators.  Our studies in Sections \ref{section:waves} and \ref{section:homoph} illustrate that the rate of reduction of seed bias is sensitive to the level of homophily and the number of sampling waves, and that the number of waves typically sampled in RDS studies may not be adequate for removing the bias induced by seed selection, especially in highly clustered populations. We also illustrate that in finite populations, it may not be wise to address seed bias by basing estimators on later waves of sampling only, as additional biases may be introduced by removing early wave samples from the possibility of inclusion in the estimator.

In our simulations (Figure \ref{cdcwave}), we illustrate that under our simulated conditions, changing from 4 to 6 waves can substantially reduce bias.
Interest in increasing the number of waves must be balanced, however, with the interest in accessing all subgroups of the target population.  The seeds are the only samples that are not directly referred by another participant.  For this reason, where it is possible to choose diverse seeds, this strategy can both reduce the variance of the estimator, as well as provide for some measure of the dependence of the final estimator on seed choice.

Our results also illustrate the dangers of high homophily (Figure \ref{cdchomoph}), increasing both the uncertainty and the effects of seed bias on the estimators.  While the homophily in the network can not be controlled by the researcher, there are some circumstances in which these results suggest RDS should not be used, or at least not used on the full population.  Clearly, when the underlying network is disconnected, the disconnected subgroups represented will be determined by the seed selection, and seed bias cannot be removed by increasing the length of sample chains.  In such populations, RDS should be used separately on each sub-population, or not at all.  
In cases of extreme homophily in a connected network, as in street-based, agency-based, and independent sex workers in Belgrade \citep{simic06}, the resulting RDS estimators will have extremely high uncertainty.  In such cases it is also advisable to use RDS either separately on each sub-population or not at all.  Where additional information from other sources can be used to estimate the relative sizes of the sub-populations, RDS estimates from separate sub-populations could be combined.

For this reason (and others), it is critical to do foundational research before conducting an RDS study.  In particular, it is important to understand the potential cleavages in the target population. Seeds should be selected to be as representative of the target population as possible, including seeds from across the various population components.  Of course, for fixed unit cost per interview, there is also a trade-off between many broadly chosen seeds and many waves of sampling.  The former allows for the diagnosis of reduction in seed bias, as well as some insurance against the case of persistent seed bias.  The latter increases the chances of actually removing such bias.

\subsection{Respondent Behavior}
Section \ref{section:bias} illustrates that additional bias can be introduced by irregularities in the sampling process.  In this simulation, we address only referral bias, however other biases may be present.  In particular, it is unclear how well respondents are able to report their numbers of alters, and how well those reports might correspond to the group within which respondents distribute their coupons.  In addition, it is not at all clear that the relationships along which coupons are passed are necessarily reciprocal.  For these reasons, it is important to improve our understanding of how RDS participants make decisions about passing coupons.

To begin to obtain this understanding, RDS researchers can add questions to their survey instruments, both the original instruments, and brief follow-ups to be administered when participants return to collect recruitment incentives or HIV test results.  Qualitative studies involving face-to-face interviews, or potentially participant observation could also be used.  These studies can begin to ascertain:

\bi
  \item How do participants choose alters to whom to pass coupons?
  \item To which alters do participants consider passing coupons?
  \item Are coupon-passing relationships likely to follow hierarchical or directed patters, or be more potentially symmetrical (or reciprocated)?
  \item Which individual characteristics might increase or decrease the likelihood of receiving or returning a coupon?
  \item Which degree-elicitation questions are most likely to correspond to the alters among whom coupons are likely to be passed?
  \item Which degree-elicitation questions are most reliably answered by the target population?
\ei
Answering these questions will be important for ensuring the reliability and validity of the results of RDS studies.  Without reliable degree-elicitation questions that reflect the size of the potential coupon-passing local network, RDS estimates will always lack a critical degree of reliability.  If coupon-distributing relationships are revealed to be non-reciprocal, more complex evaluation techniques may need to be developed.  And where referral biases can be identified and measured, evaluation techniques may take them into account to remove the resulting biases in estimation.
One approach to this type of adjustment in given in \cite{gile08} and \cite{gilehan09rds}.

\subsection{Random Walk Model}
The results of Section \ref{section:simsmall} illustrate the difficulties in using existing RDS estimators with substantial sample fractions.  When there are differential activity levels between the groups of interest (the case which RDS is designed to address), and in the presence of homophily, larger sample fractions lead to significant biases in the resulting estimators.  This bias is larger for populations with greater differentials in activity levels.
Indeed, for very large sample fractions, the sample mean out-performs the existing estimators.
In short, we find the random walk model to be an inadequate approximation to RDS sampling whenever there is a sizable sample fraction, differential activity levels by characteristic of interest, and homophily on that characteristic, circumstances which are not uncommon among populations studied via RDS.
In these cases, a new model for the sampling procedure is required.
One such model is presented in \cite{gile08} and \cite{gilehan09rds}.

\subsection{Conclusions}

We hope this paper sounds a cautionary note for the users of RDS.  Current RDS methodology is powerful and innovative.   As a sampling strategy, it effectively reaches large and varied samples of many hard-to-reach populations.  As tools for inference, the existing estimators leverage the waves of sampling to reduce the inevitable bias in convenience sampling from such populations.  On the other hand, as this study illustrates, the much-lauded allegedly {\ql}asymptotically unbiased{\qr} property of the estimators is not a panacea.  The notion of asymptotics is ambiguous and difficult in the true without-replacement case, and it is not at all clear that RDS samples reach anywhere near the depth necessary to assure any kind of convergence.  Furthermore, any convergence properties are heavily dependent on a large number of assumptions.  Practitioners should be aware of the strict conditions necessary for these desirable properties.

\bibliography{networks}

\newpage
{\bf \large APPENDIX}

\appendix{} \label{extraplots5}

In order to maintain consistent plotting ranges on all plots of estimated proportion infected, some data were outside the plot range of some earlier plots.  For completeness, those plots are repeated here, with their full ranges of data values visible. \knote{Need to add plot for with replacement example - not in thesis so need to generate.}

\newcommand{\shoe}{2in}
\newcommand{\lace}{2in}

\begin{figure}[h]
\begin{center}
\subfigure[Figure \ref{cdcwave}]
{
    \includegraphics[width=\shoe,height=\lace]{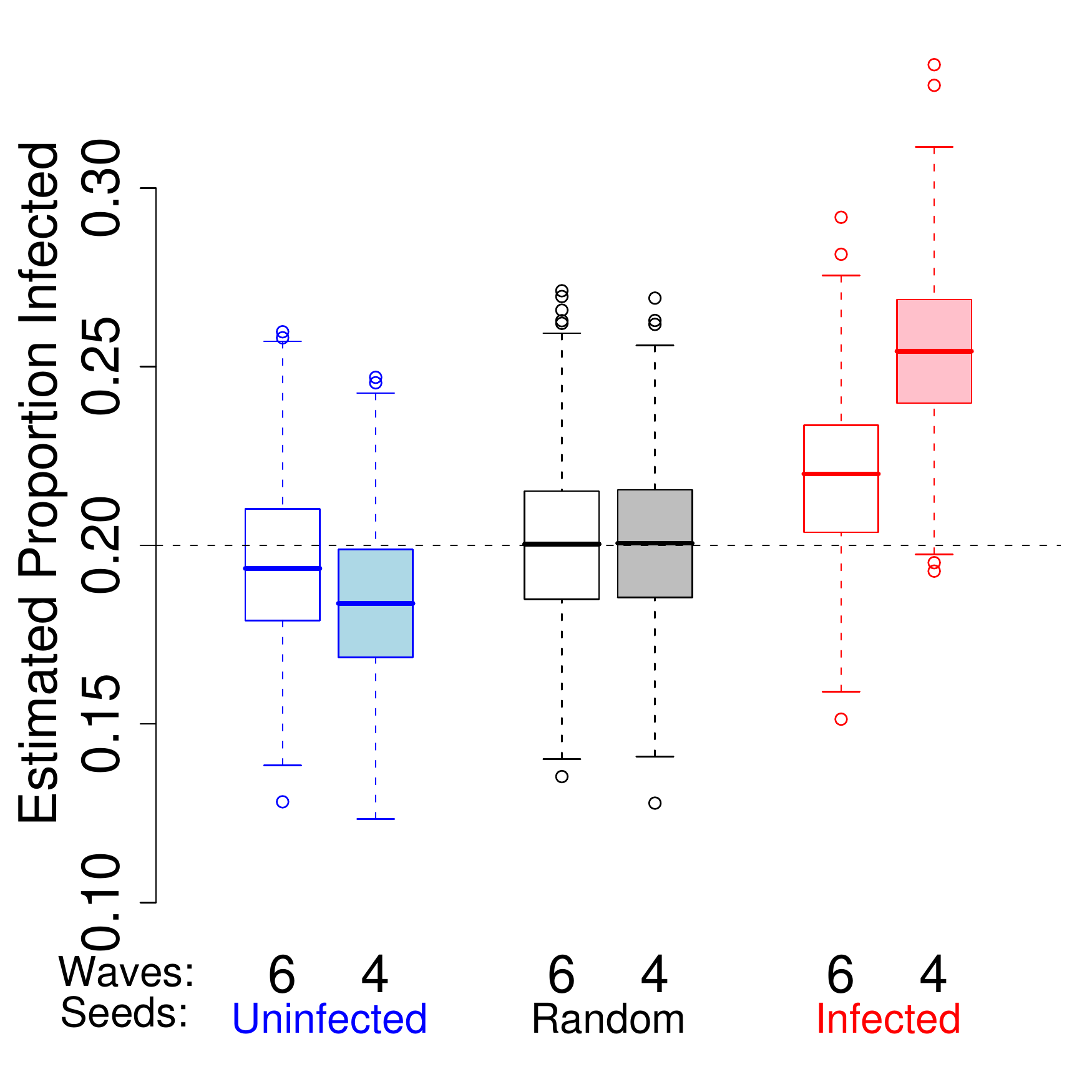}
} \hspace{.5cm}
\subfigure[Figure \ref{cdchomoph}]
{
    \includegraphics[width=\shoe,height=\lace]{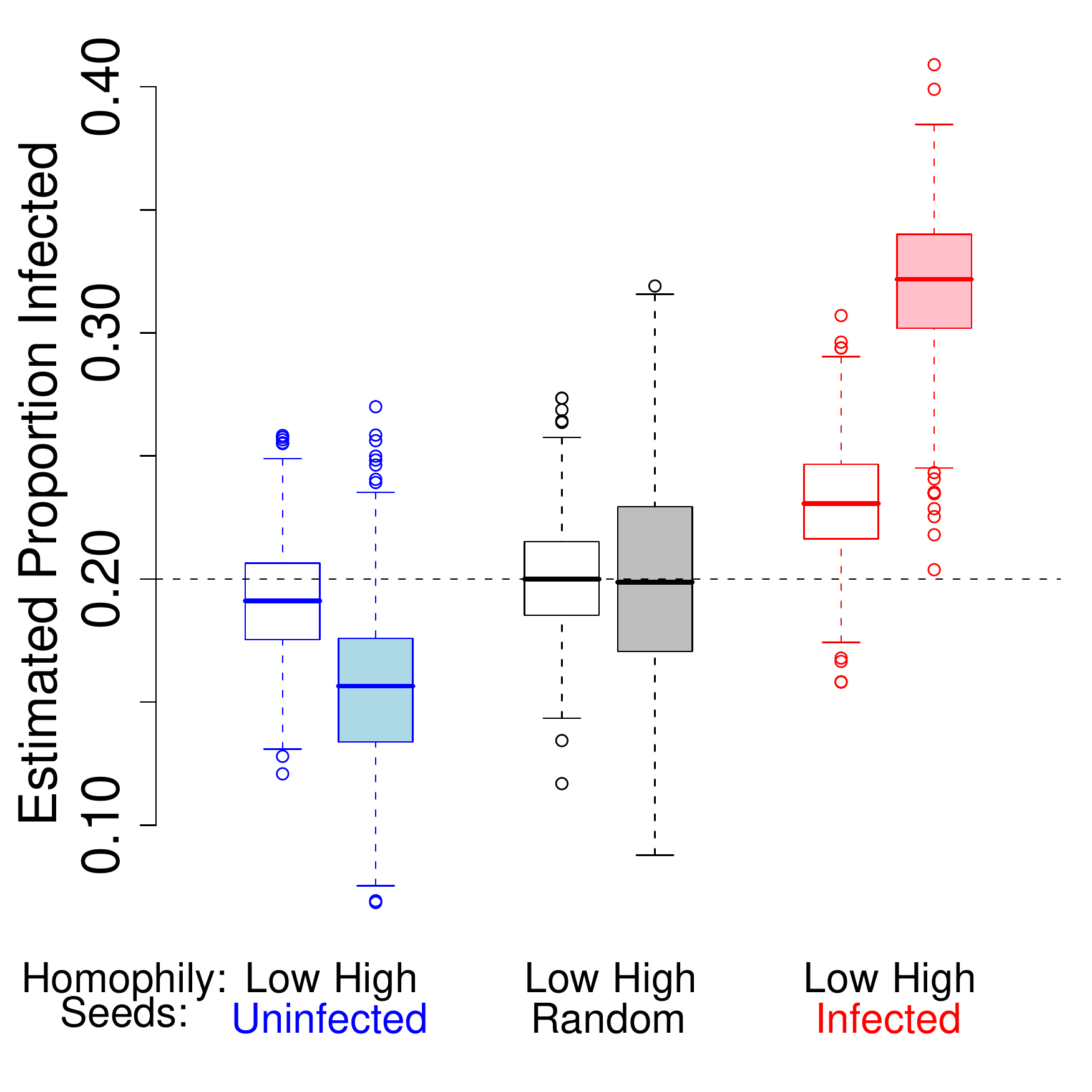}
}\hspace{.5cm}
\subfigure[Figure \ref{withouts123}]
{
    \includegraphics[width=\shoe,height=\lace]{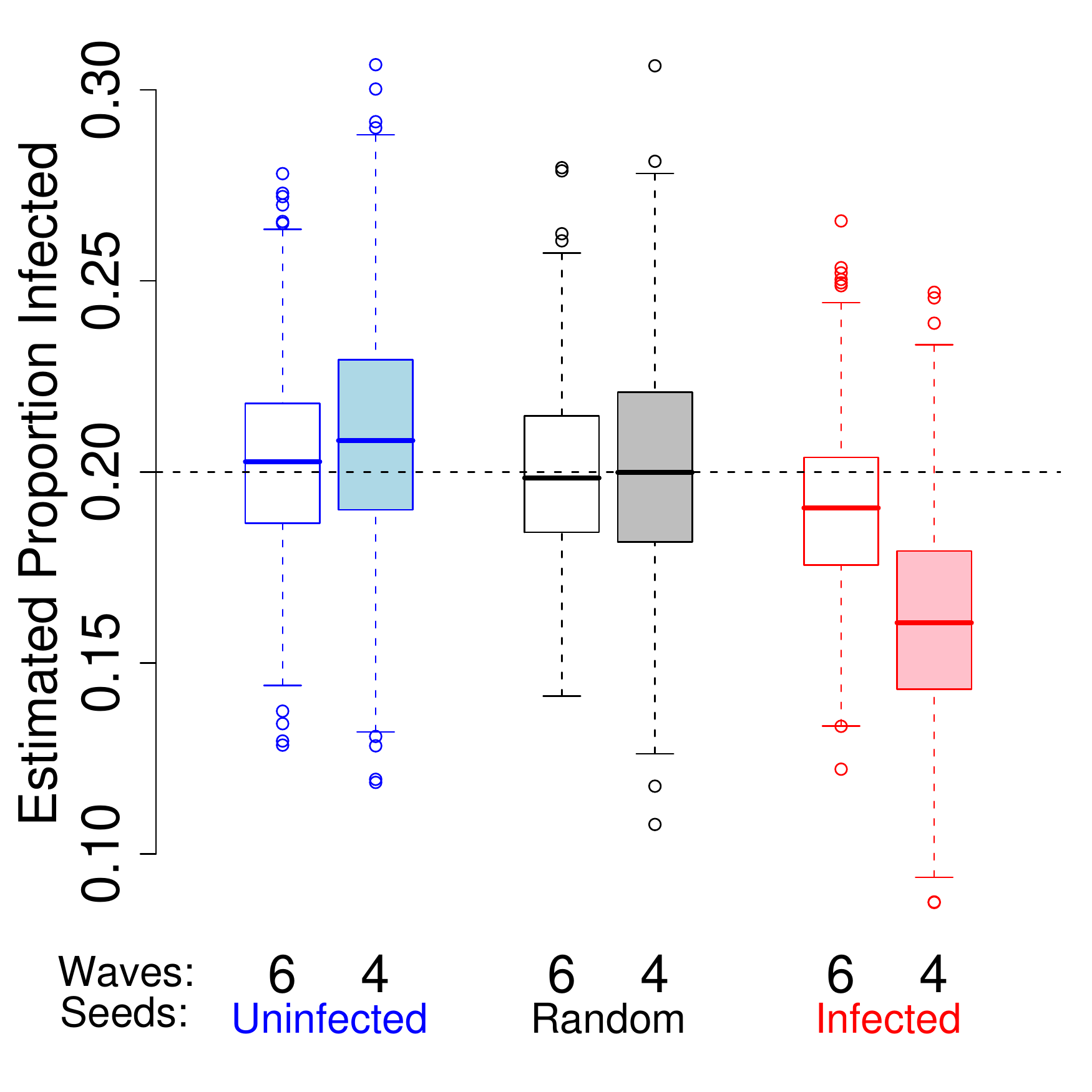}
}\vspace{.5cm}
\subfigure[Figure \ref{small4}]
{
    \includegraphics[width=\shoe,height=\lace]{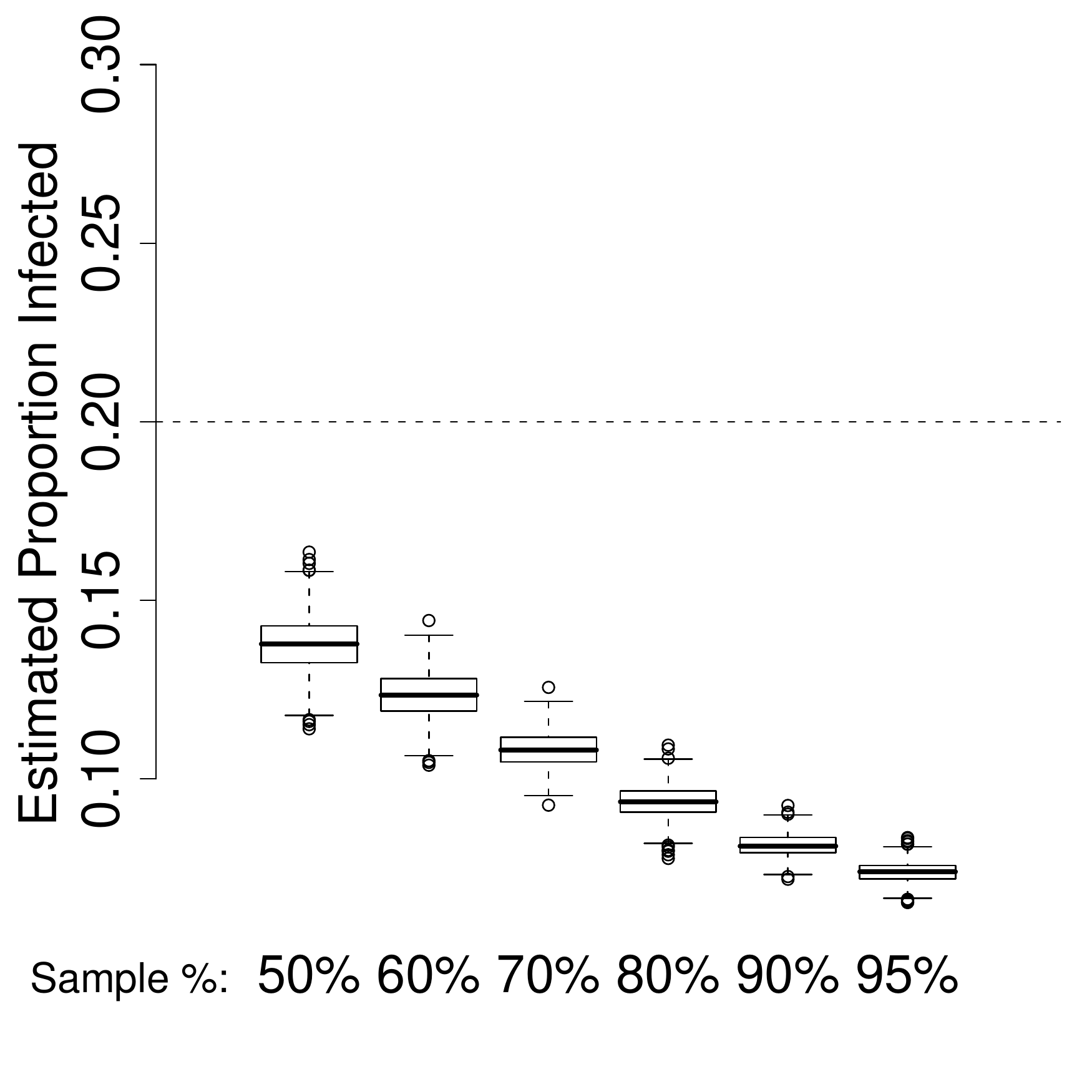}
}\end{center}  \caption{Truncated plots repeated for completeness.}
\end{figure}

\end{document}